\documentclass[%
 reprint, 
 amsmath,amssymb,
 aps,
prl,
floatfix,
]{revtex4-1}

\usepackage{graphicx}
\graphicspath{{/Users/sebastianwill/Documents/Columbia/Papers/2021-Carbon2/Tex/09-08-2021/}}
\usepackage{epsfig} 
\usepackage{dcolumn}
\usepackage{bm}
\usepackage{epstopdf}
\usepackage{multirow}
\usepackage{amsmath}
\usepackage{booktabs}
\usepackage{color}
\usepackage{gensymb}
\usepackage{kantlipsum}  
\usepackage{hyperref}
\usepackage{ulem}
\usepackage{braket}
\usepackage{physics}
\usepackage{url}
\usepackage[utf8]{inputenc}
\usepackage[T1]{fontenc}
\usepackage{mathptmx}
\usepackage{lineno}

\newcommand*{\red}{\textcolor{red}} 

\begin{document}

\title{Laser Cooling Scheme for the Carbon Dimer ($^{12}$C$_2$)}

\preprint{APS/123-QED}

\author{N.~Bigagli$^1$}
\author{D.~W.~Savin$^2$}
\author{S.~Will$^1$}
\affiliation{%
$^1$Department of Physics, Columbia University, New York, New York 10027, USA
}
\affiliation{%
$^2$Columbia Astrophysics Laboratory, Columbia University, New York, New York 10027, USA
}
\date{\today}

\begin{abstract}
We report on a scheme for laser cooling of $^{12}$C$_2$. We have calculated the branching ratios for cycling and repumping transitions and calculated the number of photon scatterings required to achieve deflection and laser cooling of a beam of $C_2$ molecules under realistic experimental conditions. Our results demonstrate that C$_2$ cooling using the Swan ($d^3\Pi_\text{g} \leftrightarrow a^3\Pi_\text{u}$) and Duck ($d^3\Pi_\text{g} \leftrightarrow c^3\Sigma_\text{u}^+$) bands is achievable via  techniques similar to state-of-the-art molecular cooling experiments. The Phillips ($A^1\Pi \red{}_\text{u} \leftrightarrow X^1\Sigma_\text{g}^+$) and Ballik-Ramsay ($b^3\Sigma_\text{g}^- \leftrightarrow a^3\Pi_\text{u}$) bands offer the potential for narrow-line cooling. This work opens up a path to cooling of molecules with carbon-carbon bonds and may pave the way toward quantum control of organic molecules. 

\end{abstract}

\maketitle

Ultracold molecules offer exciting prospects for quantum sensing \cite{carr2009cold, hudson2011improved, baron2014order}, quantum chemistry \cite{krems2009cold,hu2019direct}, quantum simulation \cite{baranov2012condensed, micheli2006toolbox, buchler2007strongly, capogrosso2010quantum, gorshkov2011tunable, moses2017new, blackmore2018ultracold}, and quantum computing \cite{demille2002quantum, park2017second, yu2019scalable}. To pursue such applications, laser cooling of molecules~\cite{FITCH2021157} has been successfully demonstrated over the recent years, utilizing molecules with quasi-diagonal Franck-Condon factors, such as SrF \cite{shuman2009radiative, barry2014magneto}, CaF \cite{zhelyazkova2014laser, hemmerling2016laser, anderegg2017radio}, YbF \cite{lim2018laser}, BaH \cite{mcnally2020optical}, CaOH \cite{baum20201d}, CaOCH$_3$ \cite{mitra2020direct}, YO \cite{yeo2015rotational}, SrOH \cite{kozyryev2017sisyphus}, and YbOH \cite{augenbraun2020laser}. These molecules are suitable for laser cooling, but their applications beyond fundamental physics  are limited. Extending laser cooling and quantum control to molecules with significant off-diagonal Franck-Condon factors, including molecules that play a broader role in chemistry, will open up intriguing research frontiers.

Molecules and compounds containing carbon atoms are of paramount importance across science. Networks of carbon atoms, such as graphene \cite{novoselov2004electric,geim2010rise}, carbon nanotubes \cite{o2018carbon}, and fullerenes \cite{dresselhaus1996science, changala2019rovibrational}, have unique electronic, optical, and mechanical properties. Molecules involving carbon-carbon and carbon-hydrogen bonds are fundamental to organic chemistry. Carbon-bearing molecules and compounds are also observed in astrophysical environments \cite{mcguire20182018} and play an important role in planet formation and astrobiology \cite{ehrenfreund2010cosmic}. 
Quantum control of carbon-bearing molecules would offer the opportunity to prepare isotopically pure molecular samples and provide a new tool set for precision studies of these molecules.

Here, we develop a laser cooling scheme for the carbon dimer (C$_2$). C$_2$ is the minimal instance of a carbon chain and a building block of organic molecules. The C-C bond is relevant in alkenes, allenes, and aromatics \cite{ouellette2018organic};  C$_2$ is used as a center for the nucleation of carbon nanotubes \cite{motaung2010situ}; and it has been observed in interstellar space \cite{souza1977detection}. Laser cooling on the C-C bond may enable laser cooling of more complex molecules with a similar bond. 
Direct optical control of atomic carbon is extremely challenging, as any relevant cooling transitions lie deep in the ultraviolet range that is difficult to access experimentally \cite{NIST_ASD, wells2011prospects}. In contrast, relevant electronic transitions of carbon dimers lie in the visible and infrared ranges that are more readily accessible.

\begin{figure}
    \centering
    \includegraphics[width = 8.6 cm]{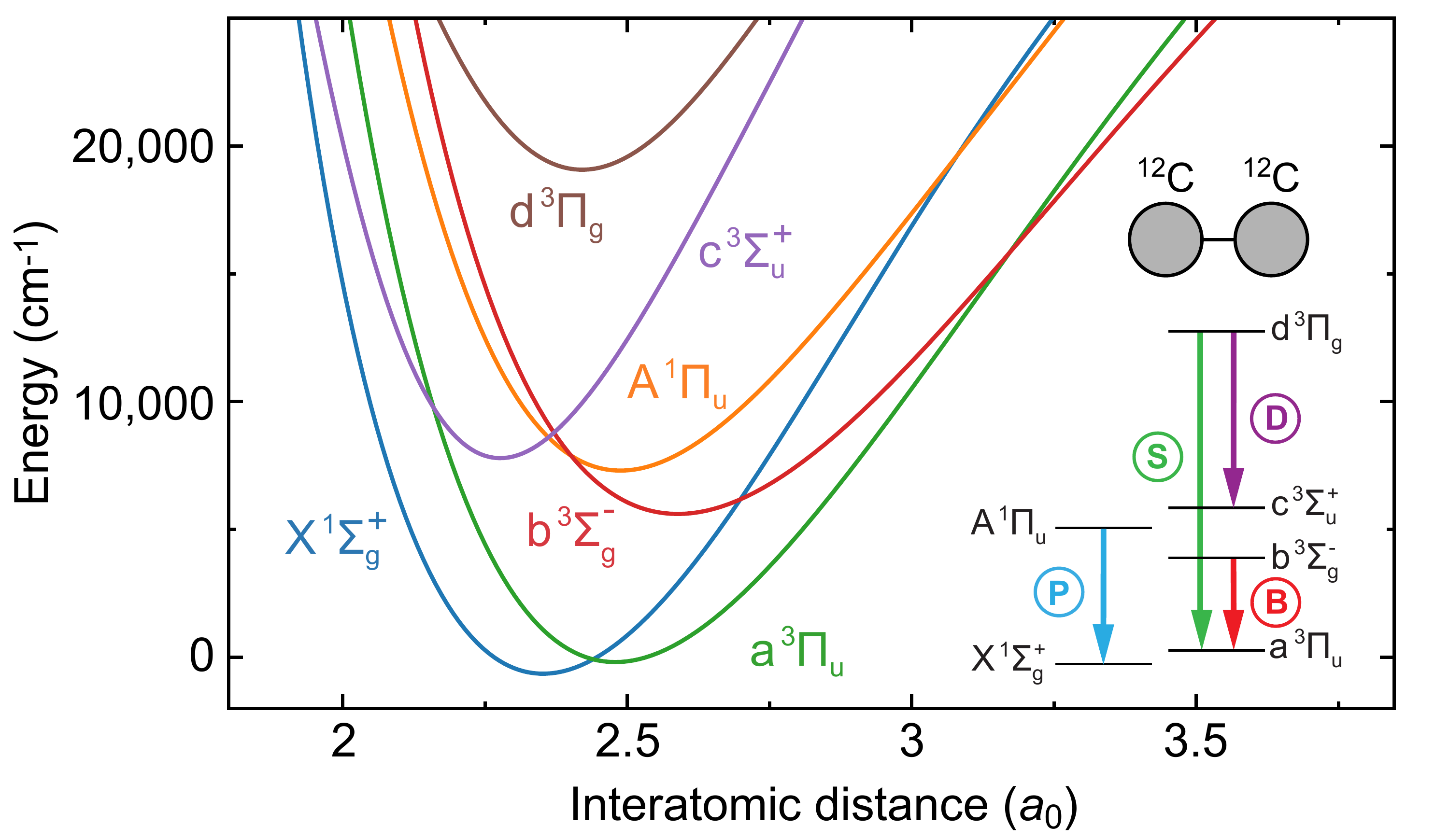}\\
    \caption{Potential energy curves of C$_2$ showing the two lowest lying singlet and the four lowest lying triplet states based on molecular constants from Refs.~\cite{chen2015simultaneous, brooke2013line, kokkin2007oscillator}. The inset shows a sketch of the C$_2$ molecule and a simplified level diagram indicating the Phillips (P, blue), Ballik-Ramsay (B, red), Swan (S, green), and Duck (D, purple) bands.}
    \label{fig:PECs}
\end{figure}

The carbon dimer has several features that are favorable for a laser cooling scheme: (1) The low-energy electronic structure features only singlet and triplet states similar to group 2 atoms and bialkali molecules. (2) The rovibrational spectrum is sparse due to relatively large vibrational and rotational spacings of about $1500\,\mathrm{ cm}^{-1}$ and $1.5\,\mathrm{ cm}^{-1}$, respectively, which limits the number of possible decay paths. This is a result of the small mass of carbon atoms and the tight bond between them (Fig.~\ref{fig:PECs}). (3) $^{12}$C, the most abundant carbon isotope with a relative abundance of $98.9\%$ \cite{emsley1995elements}, has no nuclear spin  ($I=0$). Therefore, the $^{12}$C dimer does not have a hyperfine structure. This is a remarkable simplification over the general case where a complex hyperfine structure can lead to dark states and inhibit efficient photon cycling. (4) The energy scale of spin-orbit coupling (with coupling constants on the order of $10$ cm$^{-1}$ \cite{chen2015simultaneous}) is smaller than the typical spacings between vibrational levels. Hence, population loss into the opposite spin manifold during laser cooling is expected to be small~\cite{chen2015simultaneous}. 

We analyze three laser cooling schemes. We evaluate them in terms of photon-cycling closure and calculate the achievable light forces. Specifically, we consider four transition bands between the six lowest lying electronic states of C$_2$ (Fig.~\ref{fig:PECs}). These are the Phillips ($A^1\Pi_\text{u} \leftrightarrow X^1\Sigma_\text{g}^+$), the Ballik-Ramsay ($b^3\Sigma_\text{g}^- \leftrightarrow a^3\Pi_\text{u}$), the Swan ($d^3\Pi_\text{g} \leftrightarrow a^3\Pi_\text{u}$), and the Duck ($d^3\Pi_\text{g} \leftrightarrow c^3\Sigma_\text{u}^+$) bands. Other transition bands between these electronic states do not exist due to selection rules, as detailed below. We find that the Swan and Duck bands together offer a promising pathway for laser cooling of a C$_2$ beam. Due to small scattering rates, the Phillips and Ballik-Ramsay bands only provide weak optical forces, but may be suitable for narrow-line cooling, similar to laser cooling schemes for alkaline earth atoms \cite{vogel1999, katori1999, kuwamoto1999}.

\begin{figure}
    \centering
    \includegraphics[width = 8.6 cm]{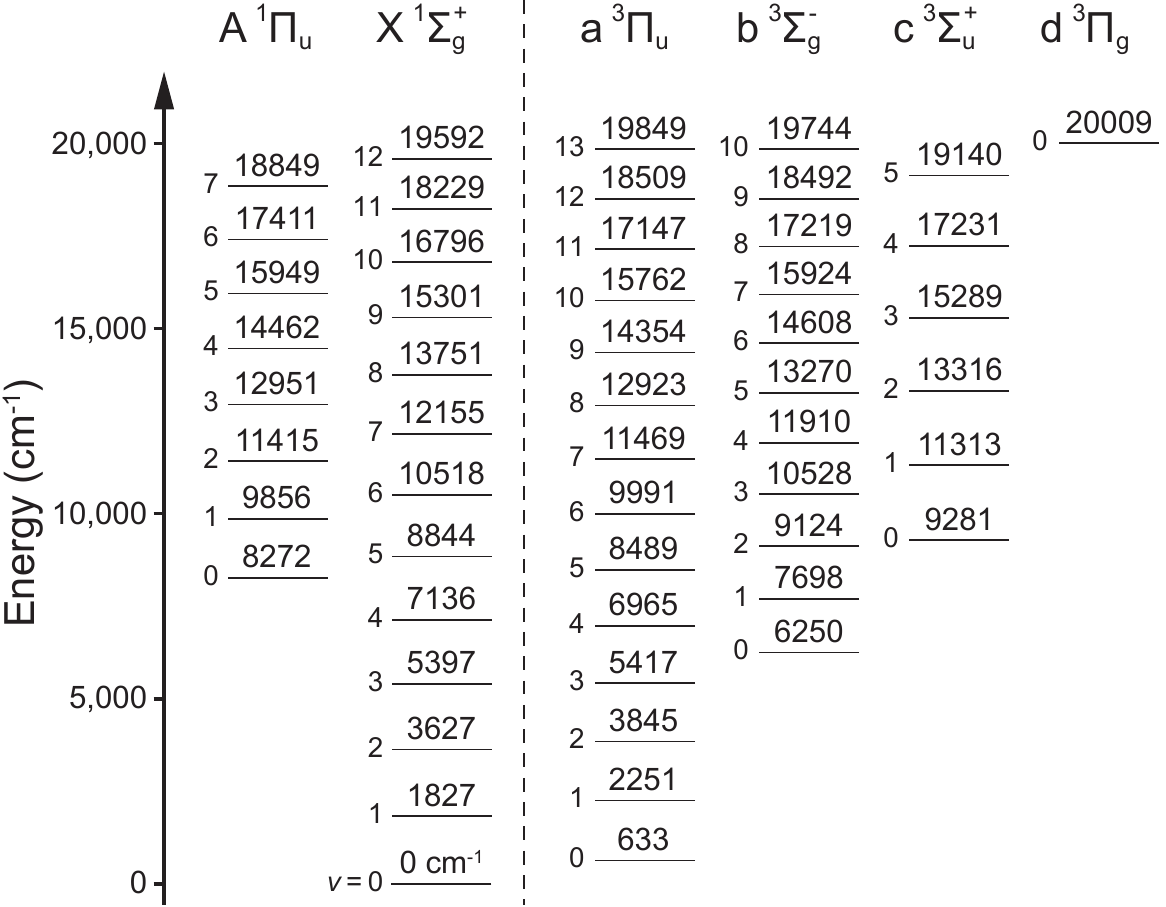}\\
    \caption{Vibrational levels of C$_2$ relevant to this study. The vibrational quantum number $v$ is shown on the left of each line. The energy values on top of the lines refer to the lowest total angular momentum quantum number, $J$, of the respective vibrational level.}
    \label{fig:C2levels}
\end{figure}

For our analysis we use existing spectroscopic data. The spectrum of C$_2$ has been extensively studied \cite{weltner1989carbon, chen2015simultaneous} to facilitate its identification in astrophysical environments \cite{brooke2013line, kokkin2007oscillator}, to determine thermodynamic functions relevant to astrochemical reactions \cite {furtenbacher2016experimental}, and to benchmark quantum chemistry calculations \cite{abrams2004full}. Our main data source is the ExoMol Project \cite{tennyson2016exomol, yurchenko2018exomol,mckemmish2020update,ExoMol}, which uses a combination of first principle theoretical calculations and experimental data. From this data, we extract transition frequencies and Einstein $A$ coefficients, from which we calculate branching ratios (BRs) of possible decay paths \cite{SI}. The branching ratios and scattering rates fully characterize the decay paths of spontaneous emissions and the  molecule-light interaction; a separate evaluation of Franck-Condon factors is not needed. An overview of the vibrational levels relevant to this study is shown in Fig.~\ref{fig:C2levels}. 


\begin{figure*}
    \centering
    \includegraphics[width = 17.2 cm]{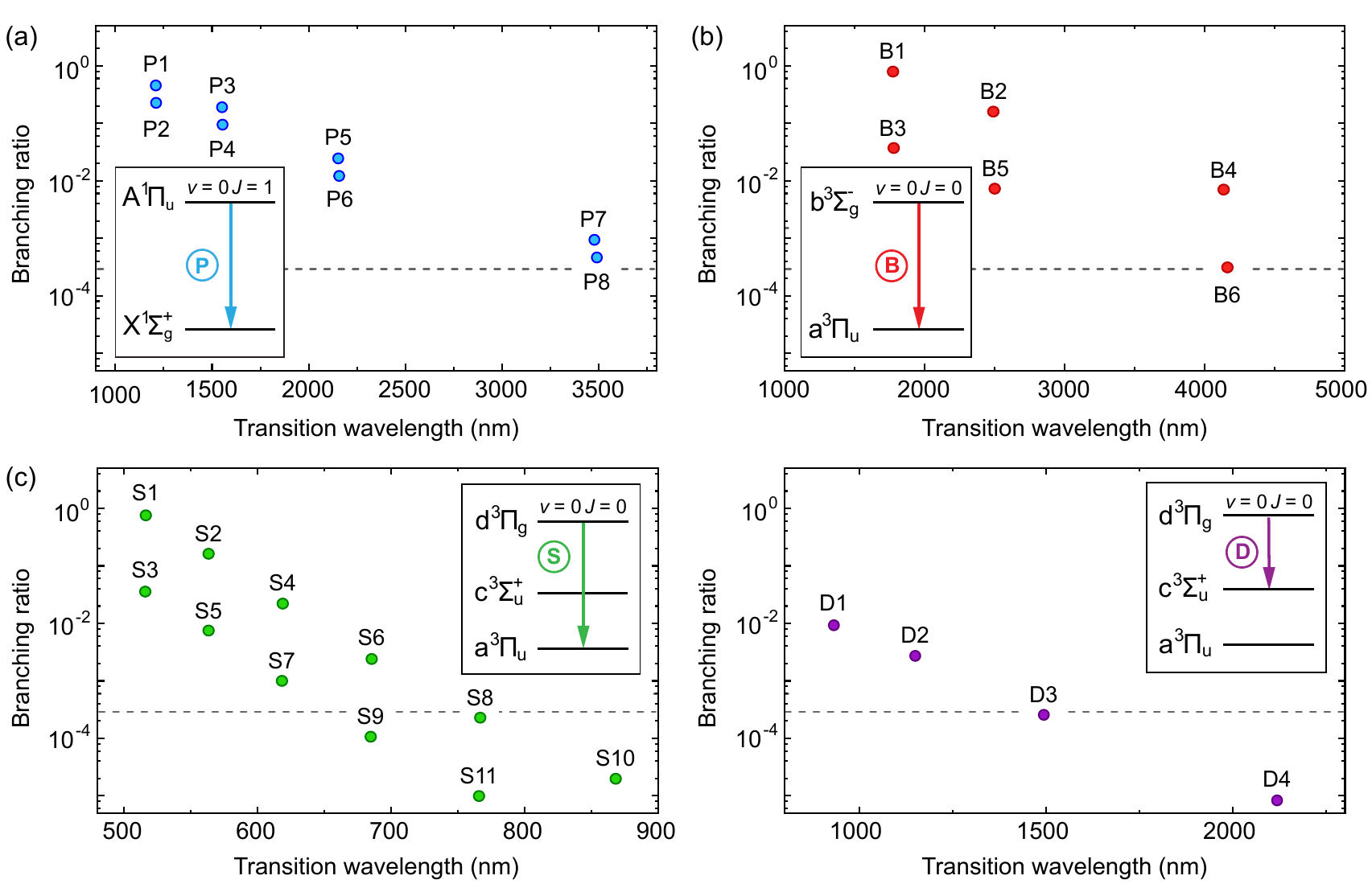}\\
    \caption{Branching ratios for decay from the excited states as specified in the main text for the (a) Phillips, (b) Ballik-Ramsay, (c) Swan and Duck bands. The decay paths for each scheme are numbered in order of decreasing branching ratios. Color codes follow the transition bands in Fig.~\ref{fig:PECs}. The horizontal dotted lines indicate a branching ratio of $3\cdot10^{-4}$, the observed cutoff to reach sufficient closure in a longitudinal cooling experiment. The labels for each transition refer to the energy levels shown in Fig.~\ref{fig:Scheme}.}
    \label{fig:BRs}
\end{figure*}

We first discuss the starting conditions for the three cycling schemes under consideration. For the Phillips band, we start in the absolute ground state, $X^1\Sigma_\text{0,g}^+$ $\ket{v,J}=\ket{0,0}$, and excite to the $A^1\Pi_\text{\,-1,u}$ $\ket{0,1}$ state. Here, $v$ is the vibrational quantum number and $J$ the total angular momentum. The subscripts 0 and -1 indicate the $\Omega$ quantum number, which is the projection of the total electronic angular momentum onto the internuclear axis. From the excited state, population decays into vibrational states of $X^1\Sigma_\text{0,g}^+$ with $J=0$ and $J=2$. The relevant branching ratios are shown in Fig.~\ref{fig:BRs}(a), calculated using the Einstein coefficients $A_{i}$ for spontaneous decay to a state $i$ \cite{yurchenko2018exomol}. Branching ratios are given by $\mathrm{BR}_i = A_{i}/\sum_j A_{j}$, where the sum runs over all possible decay paths. The splittings between the $J=0$ and $J=2$ states are about $10\, \mathrm{cm}^{-1} \approx h \times 300\,\mathrm{GHz}$ ($h$ is Planck's constant). While decays to $b^3\Sigma_\text{g}^-$ are spin forbidden, they happen with BRs $<10^{-9}$ and are therefore irrelevant for our study. Decays to $a^3\Pi_\text{u}$ are not observed, as they are spin forbidden, reflection-symmetry forbidden, and do not fulfill the $\mathit{gerade}\, \leftrightarrow \, \mathit{ungerade}$ selection rule. 

For the Ballik-Ramsay band, we start from the $a^3\Pi_{-1,\text{u}}$ $\ket{0,1}$ state and excite to the $b^3\Sigma_\text{0,g}^-$ $\ket{0,0}$ state. Excitation from $J=1$ to $J=0$ is advantageous, reducing the number of possible decay paths. The relevant BRs are shown in Fig.~\ref{fig:BRs}(b). Only decays back into $J=1$ states of $a^3\Pi_\text{u}$ are relevant, while decays to $X^1\Sigma_\text{g}^+$ are spin forbidden and strongly suppressed \cite{lefebvre2004spectra}. The $J=1$ states comprise $\Omega = 0$ and $\Omega = -1$ states that are split by about $17\, \mathrm{cm}^{-1} \approx h \times 500\,\mathrm{GHz}$. 

For the Swan and Duck scheme, we start in the metastable triplet vibrational ground state  $a^3\Pi_\text{u}$ $\ket{0,1}$ and excite to the $d^3\Pi_\text{g}$ $\ket{0,0}$ state. Transitions between $d^3\Pi_\text{g}$ $\ket{0,0}$ and $a^3\Pi_\text{u}$ $\ket{0,0}$ are strongly suppressed as $J=0\rightarrow0$ transitions are forbidden \cite{brown2003rotational}. The relevant BRs for decay into $c^3\Sigma_\text{u}^+$ and $a^3\Pi_\text{u}$ are shown in Fig.~\ref{fig:BRs}(c). The doublets for decay into $a^3\Pi_\text{u}$ have identical $v$ and $J$ and differing $\Omega$, similar to the Ballik-Ramsay band. Decay from $d^3\Pi_\text{g}$ to $A^1\Pi_u$ is not fully suppressed but extremely weak (BR $<10^{-6}$) due to the small triplet-singlet mixing and can be neglected here. Decay from $c^3\Sigma_\text{u}^+$ to $b^3\Sigma_\text{g}^-$ is suppressed as radiative transitions cannot change the $+/-$ symmetry of $\Sigma$ states \cite{brown2003rotational}. Transitions from $c^3\Sigma_\text{u}^+$ to $a^3\Pi_\text{u}$ are not allowed due to the $\mathit{gerade}\, \leftrightarrow \, \mathit{ungerade}$  selection rule \cite{brown2003rotational}. Forbidden decays to $A^1\Pi_\text{u}$ due to spin-orbit coupling have BRs $<10^{-7}$ and are therefore irrelevant to our study.


For each cycling scheme, we determine how many photons can be scattered, while retaining at least $10\%$ of the initial molecular sample in bright states \cite{SI,di2004laser}. This serves as a metric for the cycling efficiency. By driving additional repumping transitions, the closure of the cycling schemes is improved, enabling a larger number of scatterings before reaching the $10\%$-retention metric. We model photon scattering as a Bernoulli process with a probability $p$ that the molecule will remain within the cycling scheme per photon scattering \cite{di2004laser} and a probability $1-p$ that the molecule will leave. As such, $p$ is a measure of closure for a photon cycling scheme. It depends on the number of driven transitions and is given by $p = \sum_i \text{BR}_{i}$, where $i$ runs over all driven transitions. Fig.~\ref{fig:BRs} shows the BRs of the relevant decay paths out of the excited state of each cycling scheme. The fraction of molecules in bright states after $n$ cycles is given by $p^{n}$. The number of scatterings that retain $10\%$ of the molecules in a bright state is $n_{10\%} = \ln(0.1)/\ln(p)$.

For each scheme, we also determine how long it will take for a certain number of photon scatterings to happen. Combined with the initial velocity of the molecules, this determines whether cooling can be achieved within the spatial dimensions of a practical experimental setup. The time for $n_{10\%}$ scatterings is given by $t_{10\%} = n_{10\%}/R$, where $R = \Gamma/(G + 1 + 2 \sum I_{\text{sat},i}/I_i )$ is the scattering rate \cite{tarbutt2013design, FITCH2021157, SI}. Here $\Gamma$ is the excited state linewidth; $I_i$ is the intensity of the laser addressing the $i^\text{th}$ transition, set to $I_i = 10^3\,$mW$\,$cm$^{-2}$; $I_{\text{sat},i} = \pi h c \Gamma / 3 \lambda_i^3$ is the associated saturation intensity \cite{FITCH2021157}; and $G$ is the number of driven transitions. In the definition of saturation intensity, $c$ is the speed of light, and $\lambda_i$ the wavelength of the addressed transition. Saturation intensities are $\lesssim 1.5$ mW$\,$cm$^{-2}$ for the Swan and Duck bands, $\lesssim 1$ $\mu$W$\,$cm$^{-2}$ for the Phillips band and $\lesssim 0.2$ $\mu$W$\,$cm$^{-2}$ for the Ballik-Ramsay band.

We calculate $n_{10\%}$ and $t_{10\%}$ for all three schemes. The results in Tab.~\ref{tab:ClosureScatterings} show that strong closure ($p> 0.999$) can be reached by addressing 8 transitions in the Phillips scheme, 6 transitions in the Ballik-Ramsay scheme, and 9 transitions in the Swan and Duck scheme. However, the time to complete a few thousand scattering processes varies from about one second for the Phillips and Ballik-Ramsey schemes to a few milliseconds for the Swan and Duck scheme.


\begin{table}[h]
    \centering
    \caption{Closure, number of scatterings, and scattering time for an increasing number of driven transitions in the Swan and Duck (S and D), Phillips (P), and Ballik-Ramsay (B) schemes.}
    \begin{tabular}{ccccc} 
    \hline
    \hline
    & Driven transitions & $p$ & $n_{10\%}$ & $t_{10\%}$ (ms) \\ \hline
    \multirow{9}{*}{{Swan and Duck}} & S1 & 0.75638 & 8 & 0.002 \\
     & S1-S2 & 0.91873 & 27 & 0.01 \\
     & S1-S3 & 0.95440 & 49 & 0.02 \\
     & S1-S4 & 0.97646 & 96 & 0.06 \\
     & S1-S4, D1 & 0.98573 & 160 & 0.1 \\
     & S1-S5, D1 & 0.99324 & 339 & 0.3 \\
     & S1-S5, D1-D2 & 0.99596 & 568 & 0.5 \\
     & S1-S6, D1-D2 & 0.99836 & 1406 & 1.5 \\
     & S1-S7, D1-D2 & 0.99936 & 3621 & 4.2 \\\hline
     \multirow{7}{*}{{Phillips}} & P1 & 0.45290 & 2 & 0.1 \\
     & P1-P2 & 0.67861 & 5 & 0.2 \\
     & P1-P3 & 0.86793 & 16 & 0.8 \\
     & P1-P4 & 0.96210 & 59 & 3.9 \\
     & P1-P5 & 0.98650 & 169 & 13 \\
     & P1-P6 & 0.99860 & 1639 & 151 \\
     & P1-P7 & 0.99953 & 4925 & 517 \\ 
     & P1-P8 & 0.99999 & 407409 & 48140 \\ \hline
     \multirow{5}{*}{{Ballik-Ramsay}} & B1 & 0.78944 & 9 & 0.3 \\
     & B1-B2 & 0.94837 & 43 & 2.3 \\
     & B1-B3 & 0.98539 & 156 & 11 \\
     & B1-B4 & 0.99266 & 312 & 28 \\
     & B1-B5 & 0.99966 & 6788 & 738 \\
     & B1-B6 & 0.99997 & 80585 & 10224 \\
    \hline
    \hline
    \end{tabular}
    \label{tab:ClosureScatterings}
\end{table}

\begin{figure}
    \centering
    \includegraphics[width = 8 cm]{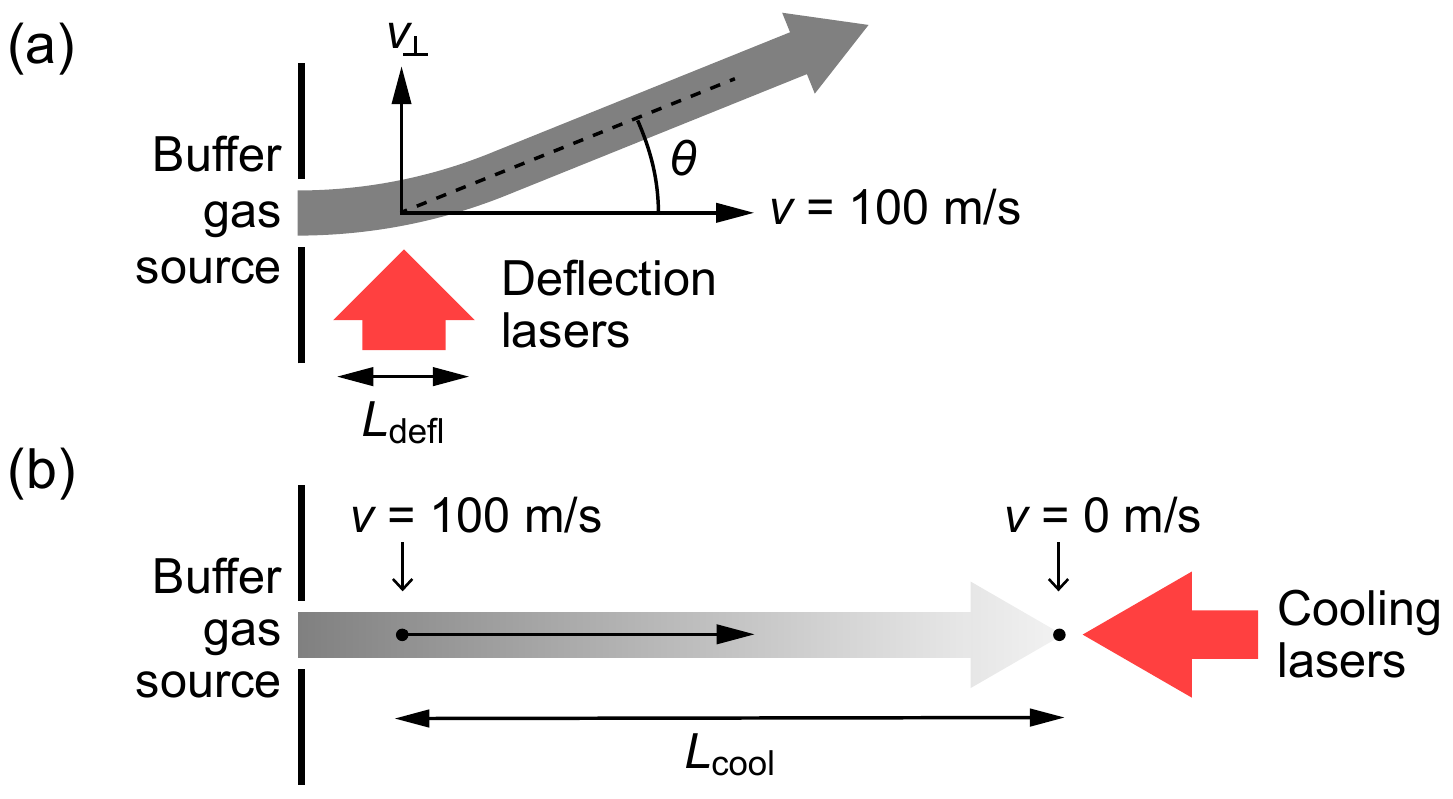}\\
    \caption{Schematic of C$_2$ (a) beam deflection and (b) longitudinal laser cooling experiments.} 
    \label{fig:Sketch}
\end{figure}

To assess the practical implications of this analysis, we consider two specific cases of molecule-light interactions for a molecular beam of C$_2$, illustrated in Fig.~\ref{fig:Sketch}: (1) transverse deflection and (2) longitudinal cooling. Deflection is a useful technique to isolate C$_2$ from a multi-species carbon beam for the preparation of pure samples. Longitudinal cooling can bring molecules to a near-standstill. For the initial molecular beam, we assume that C$_2$ emerges from a cryogenic buffer gas source with a translational temperature of $4\,$K, similar to current experiments on laser cooling of molecules \cite{maxwell2005high, shuman2009radiative, hutzler2012buffer, hemmerling2016laser, lim2018laser, mcnally2020optical}. At this temperature, the thermal forward velocity of C$_2$ is on the order of $v = 100\,$m$\,$s$^{-1}$. We note though that velocities as low as $v = 60\,$m$\,$s$^{-1}$ can be achieved in an optimized flow regime.

For transverse deflection, we assume that the lasers impinge perpendicularly on the molecular beam (Fig.~\ref{fig:Sketch}(a)). As photons are scattered, the associated recoil velocity leads to an increasing Doppler shift that detunes the molecules out of resonance with the cooling lasers. This determines the achievable deflection angle $\theta$ (see Fig.~\ref{fig:Sketch}(a)). Further photon scattering is suppressed when the Doppler shift equals the linewidth $\delta=\Gamma/2\sqrt{1+I / I_{\text{sat}}}$. For simplicity, we only take into account scattering from the main transition here and assume $I = 10^3\,$mW$\,$cm$^{-2}$. The number of scatterings $n_{\text{defl}}$ at which the Doppler detuning suppresses further photon scattering, is found by dividing the initial momentum of a molecule by that of an average photon \cite{SI}. Using $R$, we calculate the time necessary to complete $n_{\text{defl}}$ scatterings and the distance traveled by molecules in the axial direction, $L_{\text{defl}}$. From the calculated transverse velocity, $v_\perp$, and the axial velocity, $v = 100\,$m$\,$s$^{-1}$, we obtain $\theta$.

For longitudinal cooling, we assume that the laser beams counterpropagate to the molecular beam (Fig.~\ref{fig:Sketch}(b)). We also assume that the Doppler shift arising in the course of slowing is compensated by an appropriate technique, such as a counteracting Zeeman shift \cite{phillips1982laser} or chirped laser frequencies \cite{zhelyazkova2014laser}. The number of scatterings necessary to reach rest, $n_\text{cool}$, is obtained by dividing the molecule's initial momentum by that of an average photon \cite{di2004laser}. We calculate the expected distance over which the beam is slowed to a near-standstill, $L_\text{cool}$, and the time to fully slow the beam, $t_{\text{cool}}$ \cite{SI}. 

The results for deflection and cooling are shown in Table \ref{tab:Scatterings2}. The Swan and Duck scheme is favorable with a deflection angle of $\theta \sim 6^\circ$ achieved by illuminating the molecules with S1-S5 and D1 (labeled as in Fig.~\ref{fig:BRs}) over $L_\text{defl}\sim 2$ cm of axial travel, an experimentally feasible laser beam size. Cooling to a near-standstill is viable by addressing S1-S7 and D1-D2 over $L_\text{cool} \sim 16\,$cm. The deceleration due to the light force is about $3\times10^{4}\,$m$\,$s$^{-2}$ and cooling to the Doppler temperature is achieved within $\sim3\,$ms of the molecules leaving the source. The scattering rate is $R \sim 1\,$MHz. During cooling, about $3.2 \cdot 10^{3}$ photons are scattered per molecule, which corresponds to the BR cutoff of $3 \cdot 10^{-4}$ indicated in Fig.~\ref{fig:BRs}. Due to the narrow linewidth of the Phillips and Ballik-Ramsay bands (scattering rates of $R\sim 9\,$kHz), extremely long and thus impractical interaction regions are needed in a deflection experiment. Longitudinal cooling on these two bands is similarly impractical due to the small deceleration of $\sim 100\,$m$\,$s$^{-2}$ that would require a beam line of $L_\text{cool}> 10\,$m to reach standstill. Therefore, the Phillips and Ballik-Ramsay schemes are not suitable for longitudinal slowing or direct capture in a magneto-optical trap.  However, they may be useful for narrow-line cooling to reach lower Doppler temperatures in a magneto-optical trap after initial cooling. 


\begin{table}
    \centering
    \caption{Results of the photon scatterings analysis for deflection and cooling.} 
   \begin{tabular}{c|cccc} 
    \hline
    \hline
     \multirow{2}{*}{Bands} & \multicolumn{4}{c}{Transverse deflection} \\
    & $n_{\mathrm{defl}}$\ & $L_{\mathrm{defl}}$  (m)& $\theta$ ($^\circ$) & Driven transitions\\  \hline
    Swan and Duck & 338 & 0.02 & 6.1 & S1-S5, D1 \\ 
    Phillips & 611 & 5.6 & 4.4 & P1-P6  \\ 
    Ballik-Ramsay & 1941 & 21 & 9.8 & B1-B5 \\ 
    \hline
    \hline
    \multirow{2}{*}{Bands}  & \multicolumn{4}{c}{Longitudinal cooling} \\
    & $n_{\mathrm{cool}}$ & $L_\mathrm{cool}$ (m) &  $t_{\text{cool}}$ (ms) & Driven transitions\\ \hline
    Swan and Duck & $3.2\cdot 10^3$ & 0.16 & 3 & S1-S7, D1-D2 \\ 
    Phillips & $7.9\cdot 10^3$ & 47 & $9\cdot 10^2$ & P1-P8\\ 
    Ballik-Ramsay & $1.1\cdot 10^4$ & 71 & $1\cdot 10^3$ & B1-B6\\ 
    \hline
    \hline
    \end{tabular}
    \label{tab:Scatterings2}
\end{table}

\begin{figure*}
    \centering
    \includegraphics[width = 17.2 cm]{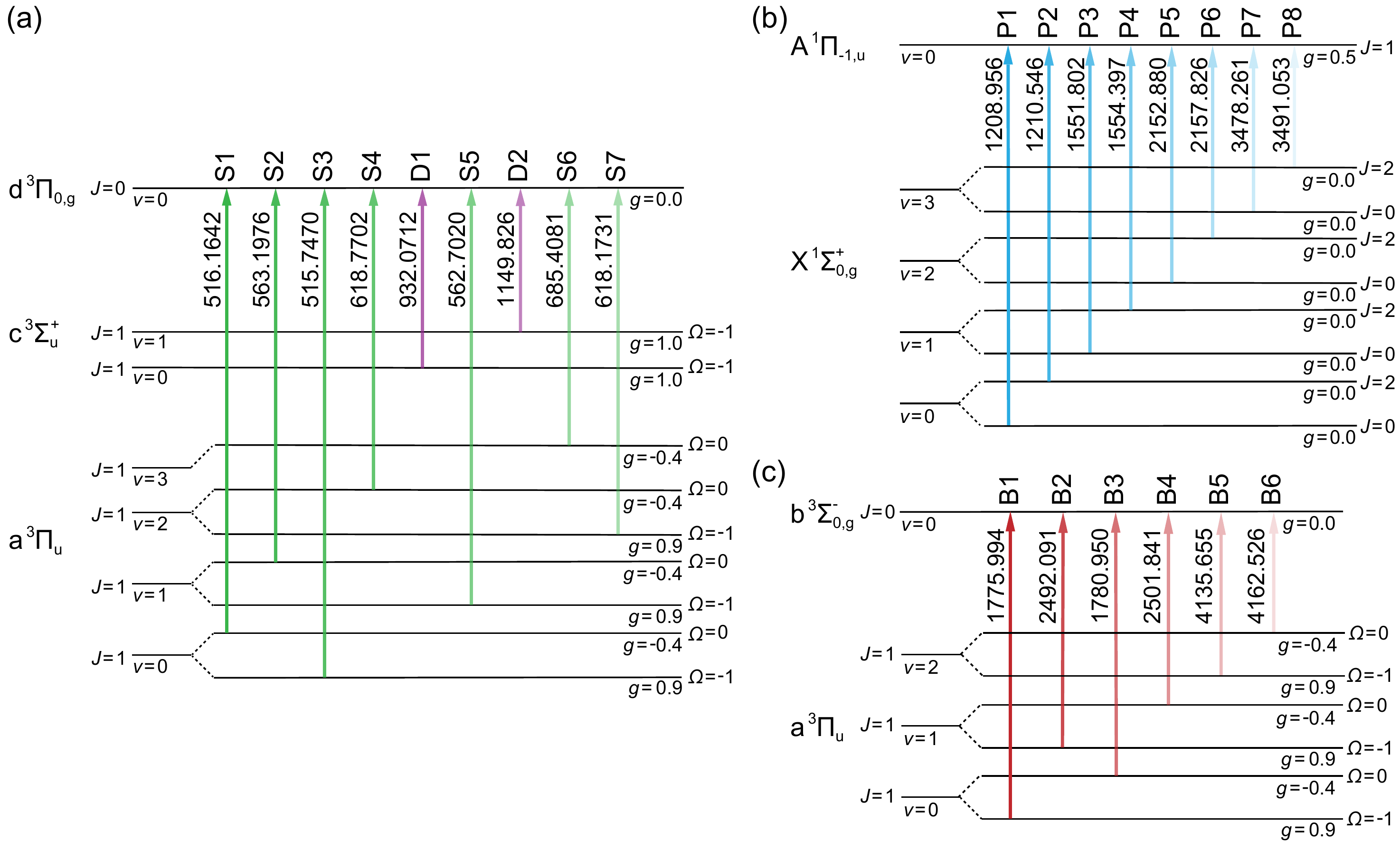}\\
    \caption{(a) Laser cooling schemes for the (a) Swan and Duck, (b) Phillips and (c) Ballik-Ramsay bands. Transitions are arranged from left to right for descending BRs. Transition wavelengths (arrow labels) are in nm. Below reach line on the right the respective Landé $g$-factor is shown. }
    \label{fig:Scheme}
\end{figure*}

Based on our findings, laser cooling of C$_2$ is conceptually feasible. Also, it should be experimentally feasible using currently available technology. Figure ~\ref{fig:Scheme}(a) shows the transition wavelengths for the Swan and Duck scheme. Laser light for all transitions can be generated with diode lasers, either directly or after frequency doubling. Diodes for 516 (S1 and S3), 685 (S6), 932 (D1, and 1150 nm (D2) are readily available. Light at 563 (S2 and S5) and 618-619 nm (S4 and S7) can be generated via frequency doubling of laser diodes. The Phillips and Ballik-Ramsay schemes are more challenging, as the infrared frequencies shown in Figs.~\ref{fig:Scheme}(b) and (c), are currently harder to generate at the required power levels.


In addition to the lasers, a cryogenic source for the generation of a cold beam of carbon beam is needed. The basic technology of cryogenic buffer gas cooling has been developed and refined over the past two decades~\cite{campbell2009cooling,hutzler2012buffer}. Hot carbon vapor to seed the buffer gas cell can be created via ablation of a graphite target~\cite{kaiser1995high,ursu2018excimer} or evaporation of graphite rods via resistive heating~\cite{savic2005reactions}. These have been shown to generate vapor containing mixtures of various carbon compounds, including C, C$_2$, C$_n$, fullerenes, etc., whose exact ratios can be controlled by varying experimental parameters. For example, ablation sources can produce carbon beams with a typical composition of $\sim10\%$ of C$_2$ \cite{kaiser1995high} and it has been shown that beam shaping of the ablation pulse can further enhance the formation of C$_2$ \cite{ursu2018excimer}. In the ablation process a significant C$_2$ fraction is found to be produced in the triplet states of $a^3\Pi_\text{u}$ \cite{mulliken1939note, ursu2018excimer}. Assuming an initial temperature of $\sim 5000 \,$K, the population in the triplet ground state is about $80\%$ of the singlet ground state \cite{krause1979carbon}. The relaxation of electron spin and rotational degrees of freedom have a cross section that is about $100$ times smaller than translational thermalization \cite{campbell2009cooling,najar2008potential,paramo2006experimental,hutzler2012buffer}. The density and in-cell time prior to extraction can be tuned such, that a significant fraction of population will be available in the $a^3\Pi_\text{u}$ $|0,1\rangle$ state for the Swan and Duck cooling scheme. Extended lifetimes in triplet electronic states of molecules with a singlet ground state have been experimentally observed previously in a buffer gas cell \cite{jadbabaie2020enhanced}.  

In conclusion, we have laid out a laser cooling scheme for C$_2$ that is experimentally feasible using state-of-the-art technology. The scheme differs from the current paradigm of laser cooling of molecules with highly diagonal Franck-Condon factors, showing that laser cooling can be possible despite significant off-diagonal Franck-Condon factors. This work lays the foundations for future studies, such as a concept for a magneto-optical trap to enable the creation ultracold ensembles of C$_2$ molecules, and possibly the trapping of individual C$_2$ molecules in optical tweezer traps \cite{anderegg2019optical}. The Landé $g$-factors shown in Fig.~\ref{fig:Scheme} suggest that a radio frequency magneto-optical trap \cite{tarbutt2015magneto} may be realized using the Swan and Duck scheme. The narrow lines in the Phillips and Ballik-Ramsey bands may be useful for coherent cooling schemes, such as bichromatic and polychromatic cooling \cite{soding1997short, chieda2011prospects, kozyryev2018coherent, wenz2020large}. More generally, our findings open up a path towards laser cooling of carbon chains and may constitute a first step towards quantum control of organic molecules.

We thank P.~C.~Stancil, I.~Stevenson, D.~Mitra, and A.~Lam for fruitful discussions and valuable comments on the manuscript. This work was supported by a Columbia University Research Initiative in Science and Engineering (RISE) award.  S.W.~acknowledges additional support from the Alfred P.~Sloan Foundation.

\bibliography{Literature}

\begin{thebibliography}{77}%
\makeatletter
\providecommand \@ifxundefined [1]{%
 \@ifx{#1\undefined}
}%
\providecommand \@ifnum [1]{%
 \ifnum #1\expandafter \@firstoftwo
 \else \expandafter \@secondoftwo
 \fi
}%
\providecommand \@ifx [1]{%
 \ifx #1\expandafter \@firstoftwo
 \else \expandafter \@secondoftwo
 \fi
}%
\providecommand \natexlab [1]{#1}%
\providecommand \enquote  [1]{``#1''}%
\providecommand \bibnamefont  [1]{#1}%
\providecommand \bibfnamefont [1]{#1}%
\providecommand \citenamefont [1]{#1}%
\providecommand \href@noop [0]{\@secondoftwo}%
\providecommand \href [0]{\begingroup \@sanitize@url \@href}%
\providecommand \@href[1]{\@@startlink{#1}\@@href}%
\providecommand \@@href[1]{\endgroup#1\@@endlink}%
\providecommand \@sanitize@url [0]{\catcode `\\12\catcode `\$12\catcode
  `\&12\catcode `\#12\catcode `\^12\catcode `\_12\catcode `\%12\relax}%
\providecommand \@@startlink[1]{}%
\providecommand \@@endlink[0]{}%
\providecommand \url  [0]{\begingroup\@sanitize@url \@url }%
\providecommand \@url [1]{\endgroup\@href {#1}{\urlprefix }}%
\providecommand \urlprefix  [0]{URL }%
\providecommand \Eprint [0]{\href }%
\providecommand \doibase [0]{http://dx.doi.org/}%
\providecommand \selectlanguage [0]{\@gobble}%
\providecommand \bibinfo  [0]{\@secondoftwo}%
\providecommand \bibfield  [0]{\@secondoftwo}%
\providecommand \translation [1]{[#1]}%
\providecommand \BibitemOpen [0]{}%
\providecommand \bibitemStop [0]{}%
\providecommand \bibitemNoStop [0]{.\EOS\space}%
\providecommand \EOS [0]{\spacefactor3000\relax}%
\providecommand \BibitemShut  [1]{\csname bibitem#1\endcsname}%
\let\auto@bib@innerbib\@empty
\bibitem [{\citenamefont {Carr}\ \emph {et~al.}(2009)\citenamefont {Carr},
  \citenamefont {DeMille}, \citenamefont {Krems},\ and\ \citenamefont
  {Ye}}]{carr2009cold}%
  \BibitemOpen
  \bibfield  {author} {\bibinfo {author} {\bibfnamefont {L.~D.}\ \bibnamefont
  {Carr}}, \bibinfo {author} {\bibfnamefont {D.}~\bibnamefont {DeMille}},
  \bibinfo {author} {\bibfnamefont {R.~V.}\ \bibnamefont {Krems}}, \ and\
  \bibinfo {author} {\bibfnamefont {J.}~\bibnamefont {Ye}},\ }\href@noop {}
  {\bibfield  {journal} {\bibinfo  {journal} {New J. Phys.}\ }\textbf {\bibinfo
  {volume} {11}},\ \bibinfo {pages} {055049} (\bibinfo {year}
  {2009})}\BibitemShut {NoStop}%
\bibitem [{\citenamefont {Hudson}\ \emph {et~al.}(2011)\citenamefont {Hudson},
  \citenamefont {Kara}, \citenamefont {Smallman}, \citenamefont {Sauer},
  \citenamefont {Tarbutt},\ and\ \citenamefont {Hinds}}]{hudson2011improved}%
  \BibitemOpen
  \bibfield  {author} {\bibinfo {author} {\bibfnamefont {J.~J.}\ \bibnamefont
  {Hudson}}, \bibinfo {author} {\bibfnamefont {D.~M.}\ \bibnamefont {Kara}},
  \bibinfo {author} {\bibfnamefont {I.}~\bibnamefont {Smallman}}, \bibinfo
  {author} {\bibfnamefont {B.~E.}\ \bibnamefont {Sauer}}, \bibinfo {author}
  {\bibfnamefont {M.~R.}\ \bibnamefont {Tarbutt}}, \ and\ \bibinfo {author}
  {\bibfnamefont {E.~A.}\ \bibnamefont {Hinds}},\ }\href@noop {} {\bibfield
  {journal} {\bibinfo  {journal} {Nature}\ }\textbf {\bibinfo {volume} {473}},\
  \bibinfo {pages} {493} (\bibinfo {year} {2011})}\BibitemShut {NoStop}%
\bibitem [{\citenamefont {Baron}\ \emph {et~al.}(2014)\citenamefont {Baron},
  \citenamefont {Campbell}, \citenamefont {DeMille}, \citenamefont {Doyle},
  \citenamefont {Gabrielse}, \citenamefont {Gurevich}, \citenamefont {Hess},
  \citenamefont {Hutzler}, \citenamefont {Kirilov}, \citenamefont {Kozyryev}
  \emph {et~al.}}]{baron2014order}%
  \BibitemOpen
  \bibfield  {author} {\bibinfo {author} {\bibfnamefont {J.}~\bibnamefont
  {Baron}}, \bibinfo {author} {\bibfnamefont {W.~C.}\ \bibnamefont {Campbell}},
  \bibinfo {author} {\bibfnamefont {D.}~\bibnamefont {DeMille}}, \bibinfo
  {author} {\bibfnamefont {J.~M.}\ \bibnamefont {Doyle}}, \bibinfo {author}
  {\bibfnamefont {G.}~\bibnamefont {Gabrielse}}, \bibinfo {author}
  {\bibfnamefont {Y.~V.}\ \bibnamefont {Gurevich}}, \bibinfo {author}
  {\bibfnamefont {P.~W.}\ \bibnamefont {Hess}}, \bibinfo {author}
  {\bibfnamefont {N.~R.}\ \bibnamefont {Hutzler}}, \bibinfo {author}
  {\bibfnamefont {E.}~\bibnamefont {Kirilov}}, \bibinfo {author} {\bibfnamefont
  {I.}~\bibnamefont {Kozyryev}},  \emph {et~al.},\ }\href@noop {} {\bibfield
  {journal} {\bibinfo  {journal} {Science}\ }\textbf {\bibinfo {volume}
  {343}},\ \bibinfo {pages} {269} (\bibinfo {year} {2014})}\BibitemShut
  {NoStop}%
\bibitem [{\citenamefont {Krems}\ \emph {et~al.}(2009)\citenamefont {Krems},
  \citenamefont {Friedrich},\ and\ \citenamefont {Stwalley}}]{krems2009cold}%
  \BibitemOpen
  \bibfield  {author} {\bibinfo {author} {\bibfnamefont {R.}~\bibnamefont
  {Krems}}, \bibinfo {author} {\bibfnamefont {B.}~\bibnamefont {Friedrich}}, \
  and\ \bibinfo {author} {\bibfnamefont {W.~C.}\ \bibnamefont {Stwalley}},\
  }\href@noop {} {\emph {\bibinfo {title} {Cold molecules: theory, experiment,
  applications}}}\ (\bibinfo  {publisher} {CRC press},\ \bibinfo {year}
  {2009})\BibitemShut {NoStop}%
\bibitem [{\citenamefont {Hu}\ \emph {et~al.}(2019)\citenamefont {Hu},
  \citenamefont {Liu}, \citenamefont {Grimes}, \citenamefont {Lin},
  \citenamefont {Gheorghe}, \citenamefont {Vexiau}, \citenamefont
  {Bouloufa-Maafa}, \citenamefont {Dulieu}, \citenamefont {Rosenband},\ and\
  \citenamefont {Ni}}]{hu2019direct}%
  \BibitemOpen
  \bibfield  {author} {\bibinfo {author} {\bibfnamefont {M.-G.}\ \bibnamefont
  {Hu}}, \bibinfo {author} {\bibfnamefont {Y.}~\bibnamefont {Liu}}, \bibinfo
  {author} {\bibfnamefont {D.~D.}\ \bibnamefont {Grimes}}, \bibinfo {author}
  {\bibfnamefont {Y.-W.}\ \bibnamefont {Lin}}, \bibinfo {author} {\bibfnamefont
  {A.~H.}\ \bibnamefont {Gheorghe}}, \bibinfo {author} {\bibfnamefont
  {R.}~\bibnamefont {Vexiau}}, \bibinfo {author} {\bibfnamefont
  {N.}~\bibnamefont {Bouloufa-Maafa}}, \bibinfo {author} {\bibfnamefont
  {O.}~\bibnamefont {Dulieu}}, \bibinfo {author} {\bibfnamefont
  {T.}~\bibnamefont {Rosenband}}, \ and\ \bibinfo {author} {\bibfnamefont
  {K.-K.}\ \bibnamefont {Ni}},\ }\href@noop {} {\bibfield  {journal} {\bibinfo
  {journal} {Science}\ }\textbf {\bibinfo {volume} {366}},\ \bibinfo {pages}
  {1111} (\bibinfo {year} {2019})}\BibitemShut {NoStop}%
\bibitem [{\citenamefont {Baranov}\ \emph {et~al.}(2012)\citenamefont
  {Baranov}, \citenamefont {Dalmonte}, \citenamefont {Pupillo},\ and\
  \citenamefont {Zoller}}]{baranov2012condensed}%
  \BibitemOpen
  \bibfield  {author} {\bibinfo {author} {\bibfnamefont {M.~A.}\ \bibnamefont
  {Baranov}}, \bibinfo {author} {\bibfnamefont {M.}~\bibnamefont {Dalmonte}},
  \bibinfo {author} {\bibfnamefont {G.}~\bibnamefont {Pupillo}}, \ and\
  \bibinfo {author} {\bibfnamefont {P.}~\bibnamefont {Zoller}},\ }\href@noop {}
  {\bibfield  {journal} {\bibinfo  {journal} {Chem. Rev.}\ }\textbf {\bibinfo
  {volume} {112}},\ \bibinfo {pages} {5012} (\bibinfo {year}
  {2012})}\BibitemShut {NoStop}%
\bibitem [{\citenamefont {Micheli}\ \emph {et~al.}(2006)\citenamefont
  {Micheli}, \citenamefont {Brennen},\ and\ \citenamefont
  {Zoller}}]{micheli2006toolbox}%
  \BibitemOpen
  \bibfield  {author} {\bibinfo {author} {\bibfnamefont {A.}~\bibnamefont
  {Micheli}}, \bibinfo {author} {\bibfnamefont {G.}~\bibnamefont {Brennen}}, \
  and\ \bibinfo {author} {\bibfnamefont {P.}~\bibnamefont {Zoller}},\
  }\href@noop {} {\bibfield  {journal} {\bibinfo  {journal} {Nat. Phys.}\
  }\textbf {\bibinfo {volume} {2}},\ \bibinfo {pages} {341} (\bibinfo {year}
  {2006})}\BibitemShut {NoStop}%
\bibitem [{\citenamefont {B{\"u}chler}\ \emph {et~al.}(2007)\citenamefont
  {B{\"u}chler}, \citenamefont {Demler}, \citenamefont {Lukin}, \citenamefont
  {Micheli}, \citenamefont {Prokof’Ev}, \citenamefont {Pupillo},\ and\
  \citenamefont {Zoller}}]{buchler2007strongly}%
  \BibitemOpen
  \bibfield  {author} {\bibinfo {author} {\bibfnamefont {H.~P.}\ \bibnamefont
  {B{\"u}chler}}, \bibinfo {author} {\bibfnamefont {E.}~\bibnamefont {Demler}},
  \bibinfo {author} {\bibfnamefont {M.}~\bibnamefont {Lukin}}, \bibinfo
  {author} {\bibfnamefont {A.}~\bibnamefont {Micheli}}, \bibinfo {author}
  {\bibfnamefont {N.}~\bibnamefont {Prokof’Ev}}, \bibinfo {author}
  {\bibfnamefont {G.}~\bibnamefont {Pupillo}}, \ and\ \bibinfo {author}
  {\bibfnamefont {P.}~\bibnamefont {Zoller}},\ }\href@noop {} {\bibfield
  {journal} {\bibinfo  {journal} {Phys. Rev. Lett.}\ }\textbf {\bibinfo
  {volume} {98}},\ \bibinfo {pages} {060404} (\bibinfo {year}
  {2007})}\BibitemShut {NoStop}%
\bibitem [{\citenamefont {Capogrosso-Sansone}\ \emph
  {et~al.}(2010)\citenamefont {Capogrosso-Sansone}, \citenamefont {Trefzger},
  \citenamefont {Lewenstein}, \citenamefont {Zoller},\ and\ \citenamefont
  {Pupillo}}]{capogrosso2010quantum}%
  \BibitemOpen
  \bibfield  {author} {\bibinfo {author} {\bibfnamefont {B.}~\bibnamefont
  {Capogrosso-Sansone}}, \bibinfo {author} {\bibfnamefont {C.}~\bibnamefont
  {Trefzger}}, \bibinfo {author} {\bibfnamefont {M.}~\bibnamefont
  {Lewenstein}}, \bibinfo {author} {\bibfnamefont {P.}~\bibnamefont {Zoller}},
  \ and\ \bibinfo {author} {\bibfnamefont {G.}~\bibnamefont {Pupillo}},\
  }\href@noop {} {\bibfield  {journal} {\bibinfo  {journal} {Phys. Rev. Lett.}\
  }\textbf {\bibinfo {volume} {104}},\ \bibinfo {pages} {125301} (\bibinfo
  {year} {2010})}\BibitemShut {NoStop}%
\bibitem [{\citenamefont {Gorshkov}\ \emph {et~al.}(2011)\citenamefont
  {Gorshkov}, \citenamefont {Manmana}, \citenamefont {Chen}, \citenamefont
  {Ye}, \citenamefont {Demler}, \citenamefont {Lukin},\ and\ \citenamefont
  {Rey}}]{gorshkov2011tunable}%
  \BibitemOpen
  \bibfield  {author} {\bibinfo {author} {\bibfnamefont {A.~V.}\ \bibnamefont
  {Gorshkov}}, \bibinfo {author} {\bibfnamefont {S.~R.}\ \bibnamefont
  {Manmana}}, \bibinfo {author} {\bibfnamefont {G.}~\bibnamefont {Chen}},
  \bibinfo {author} {\bibfnamefont {J.}~\bibnamefont {Ye}}, \bibinfo {author}
  {\bibfnamefont {E.}~\bibnamefont {Demler}}, \bibinfo {author} {\bibfnamefont
  {M.~D.}\ \bibnamefont {Lukin}}, \ and\ \bibinfo {author} {\bibfnamefont
  {A.~M.}\ \bibnamefont {Rey}},\ }\href@noop {} {\bibfield  {journal} {\bibinfo
   {journal} {Phys. Rev. Lett.}\ }\textbf {\bibinfo {volume} {107}},\ \bibinfo
  {pages} {115301} (\bibinfo {year} {2011})}\BibitemShut {NoStop}%
\bibitem [{\citenamefont {Moses}\ \emph {et~al.}(2017)\citenamefont {Moses},
  \citenamefont {Covey}, \citenamefont {Miecnikowski}, \citenamefont {Jin},\
  and\ \citenamefont {Ye}}]{moses2017new}%
  \BibitemOpen
  \bibfield  {author} {\bibinfo {author} {\bibfnamefont {S.~A.}\ \bibnamefont
  {Moses}}, \bibinfo {author} {\bibfnamefont {J.~P.}\ \bibnamefont {Covey}},
  \bibinfo {author} {\bibfnamefont {M.~T.}\ \bibnamefont {Miecnikowski}},
  \bibinfo {author} {\bibfnamefont {D.~S.}\ \bibnamefont {Jin}}, \ and\
  \bibinfo {author} {\bibfnamefont {J.}~\bibnamefont {Ye}},\ }\href@noop {}
  {\bibfield  {journal} {\bibinfo  {journal} {Nat. Phys.}\ }\textbf {\bibinfo
  {volume} {13}},\ \bibinfo {pages} {13} (\bibinfo {year} {2017})}\BibitemShut
  {NoStop}%
\bibitem [{\citenamefont {Blackmore}\ \emph {et~al.}(2018)\citenamefont
  {Blackmore}, \citenamefont {Caldwell}, \citenamefont {Gregory}, \citenamefont
  {Bridge}, \citenamefont {Sawant}, \citenamefont {Aldegunde}, \citenamefont
  {Mur-Petit}, \citenamefont {Jaksch}, \citenamefont {Hutson}, \citenamefont
  {Sauer} \emph {et~al.}}]{blackmore2018ultracold}%
  \BibitemOpen
  \bibfield  {author} {\bibinfo {author} {\bibfnamefont {J.~A.}\ \bibnamefont
  {Blackmore}}, \bibinfo {author} {\bibfnamefont {L.}~\bibnamefont {Caldwell}},
  \bibinfo {author} {\bibfnamefont {P.~D.}\ \bibnamefont {Gregory}}, \bibinfo
  {author} {\bibfnamefont {E.~M.}\ \bibnamefont {Bridge}}, \bibinfo {author}
  {\bibfnamefont {R.}~\bibnamefont {Sawant}}, \bibinfo {author} {\bibfnamefont
  {J.}~\bibnamefont {Aldegunde}}, \bibinfo {author} {\bibfnamefont
  {J.}~\bibnamefont {Mur-Petit}}, \bibinfo {author} {\bibfnamefont
  {D.}~\bibnamefont {Jaksch}}, \bibinfo {author} {\bibfnamefont {J.~M.}\
  \bibnamefont {Hutson}}, \bibinfo {author} {\bibfnamefont {B.}~\bibnamefont
  {Sauer}},  \emph {et~al.},\ }\href@noop {} {\bibfield  {journal} {\bibinfo
  {journal} {Quantum Sci. Technol.}\ }\textbf {\bibinfo {volume} {4}},\
  \bibinfo {pages} {014010} (\bibinfo {year} {2018})}\BibitemShut {NoStop}%
\bibitem [{\citenamefont {DeMille}(2002)}]{demille2002quantum}%
  \BibitemOpen
  \bibfield  {author} {\bibinfo {author} {\bibfnamefont {D.}~\bibnamefont
  {DeMille}},\ }\href@noop {} {\bibfield  {journal} {\bibinfo  {journal} {Phys.
  Rev. Let.}\ }\textbf {\bibinfo {volume} {88}},\ \bibinfo {pages} {067901}
  (\bibinfo {year} {2002})}\BibitemShut {NoStop}%
\bibitem [{\citenamefont {Park}\ \emph {et~al.}(2017)\citenamefont {Park},
  \citenamefont {Yan}, \citenamefont {Loh}, \citenamefont {Will},\ and\
  \citenamefont {Zwierlein}}]{park2017second}%
  \BibitemOpen
  \bibfield  {author} {\bibinfo {author} {\bibfnamefont {J.~W.}\ \bibnamefont
  {Park}}, \bibinfo {author} {\bibfnamefont {Z.~Z.}\ \bibnamefont {Yan}},
  \bibinfo {author} {\bibfnamefont {H.}~\bibnamefont {Loh}}, \bibinfo {author}
  {\bibfnamefont {S.~A.}\ \bibnamefont {Will}}, \ and\ \bibinfo {author}
  {\bibfnamefont {M.~W.}\ \bibnamefont {Zwierlein}},\ }\href@noop {} {\bibfield
   {journal} {\bibinfo  {journal} {Science}\ }\textbf {\bibinfo {volume}
  {357}},\ \bibinfo {pages} {372} (\bibinfo {year} {2017})}\BibitemShut
  {NoStop}%
\bibitem [{\citenamefont {Yu}\ \emph {et~al.}(2019)\citenamefont {Yu},
  \citenamefont {Cheuk}, \citenamefont {Kozyryev},\ and\ \citenamefont
  {Doyle}}]{yu2019scalable}%
  \BibitemOpen
  \bibfield  {author} {\bibinfo {author} {\bibfnamefont {P.}~\bibnamefont
  {Yu}}, \bibinfo {author} {\bibfnamefont {L.~W.}\ \bibnamefont {Cheuk}},
  \bibinfo {author} {\bibfnamefont {I.}~\bibnamefont {Kozyryev}}, \ and\
  \bibinfo {author} {\bibfnamefont {J.~M.}\ \bibnamefont {Doyle}},\ }\href@noop
  {} {\bibfield  {journal} {\bibinfo  {journal} {New J. Phys.}\ }\textbf
  {\bibinfo {volume} {21}},\ \bibinfo {pages} {093049} (\bibinfo {year}
  {2019})}\BibitemShut {NoStop}%
\bibitem [{\citenamefont {Fitch}\ and\ \citenamefont
  {Tarbutt}(2021)}]{FITCH2021157}%
  \BibitemOpen
  \bibfield  {author} {\bibinfo {author} {\bibfnamefont {N.}~\bibnamefont
  {Fitch}}\ and\ \bibinfo {author} {\bibfnamefont {M.}~\bibnamefont {Tarbutt}}\
  }(\bibinfo  {publisher} {Academic Press},\ \bibinfo {year} {2021})\ pp.\
  \bibinfo {pages} {157--262}\BibitemShut {NoStop}%
\bibitem [{\citenamefont {Shuman}\ \emph {et~al.}(2009)\citenamefont {Shuman},
  \citenamefont {Barry}, \citenamefont {Glenn},\ and\ \citenamefont
  {DeMille}}]{shuman2009radiative}%
  \BibitemOpen
  \bibfield  {author} {\bibinfo {author} {\bibfnamefont {E.}~\bibnamefont
  {Shuman}}, \bibinfo {author} {\bibfnamefont {J.}~\bibnamefont {Barry}},
  \bibinfo {author} {\bibfnamefont {D.}~\bibnamefont {Glenn}}, \ and\ \bibinfo
  {author} {\bibfnamefont {D.}~\bibnamefont {DeMille}},\ }\href@noop {}
  {\bibfield  {journal} {\bibinfo  {journal} {Phys. rev. lett.}\ }\textbf
  {\bibinfo {volume} {103}},\ \bibinfo {pages} {223001} (\bibinfo {year}
  {2009})}\BibitemShut {NoStop}%
\bibitem [{\citenamefont {Barry}\ \emph {et~al.}(2014)\citenamefont {Barry},
  \citenamefont {McCarron}, \citenamefont {Norrgard}, \citenamefont
  {Steinecker},\ and\ \citenamefont {DeMille}}]{barry2014magneto}%
  \BibitemOpen
  \bibfield  {author} {\bibinfo {author} {\bibfnamefont {J.}~\bibnamefont
  {Barry}}, \bibinfo {author} {\bibfnamefont {D.}~\bibnamefont {McCarron}},
  \bibinfo {author} {\bibfnamefont {E.}~\bibnamefont {Norrgard}}, \bibinfo
  {author} {\bibfnamefont {M.}~\bibnamefont {Steinecker}}, \ and\ \bibinfo
  {author} {\bibfnamefont {D.}~\bibnamefont {DeMille}},\ }\href@noop {}
  {\bibfield  {journal} {\bibinfo  {journal} {Nature}\ }\textbf {\bibinfo
  {volume} {512}},\ \bibinfo {pages} {286} (\bibinfo {year}
  {2014})}\BibitemShut {NoStop}%
\bibitem [{\citenamefont {Zhelyazkova}\ \emph {et~al.}(2014)\citenamefont
  {Zhelyazkova}, \citenamefont {Cournol}, \citenamefont {Wall}, \citenamefont
  {Matsushima}, \citenamefont {Hudson}, \citenamefont {Hinds}, \citenamefont
  {Tarbutt},\ and\ \citenamefont {Sauer}}]{zhelyazkova2014laser}%
  \BibitemOpen
  \bibfield  {author} {\bibinfo {author} {\bibfnamefont {V.}~\bibnamefont
  {Zhelyazkova}}, \bibinfo {author} {\bibfnamefont {A.}~\bibnamefont
  {Cournol}}, \bibinfo {author} {\bibfnamefont {T.~E.}\ \bibnamefont {Wall}},
  \bibinfo {author} {\bibfnamefont {A.}~\bibnamefont {Matsushima}}, \bibinfo
  {author} {\bibfnamefont {J.~J.}\ \bibnamefont {Hudson}}, \bibinfo {author}
  {\bibfnamefont {E.}~\bibnamefont {Hinds}}, \bibinfo {author} {\bibfnamefont
  {M.}~\bibnamefont {Tarbutt}}, \ and\ \bibinfo {author} {\bibfnamefont
  {B.}~\bibnamefont {Sauer}},\ }\href@noop {} {\bibfield  {journal} {\bibinfo
  {journal} {Phys. Rev. A}\ }\textbf {\bibinfo {volume} {89}},\ \bibinfo
  {pages} {053416} (\bibinfo {year} {2014})}\BibitemShut {NoStop}%
\bibitem [{\citenamefont {Hemmerling}\ \emph {et~al.}(2016)\citenamefont
  {Hemmerling}, \citenamefont {Chae}, \citenamefont {Ravi}, \citenamefont
  {Anderegg}, \citenamefont {Drayna}, \citenamefont {Hutzler}, \citenamefont
  {Collopy}, \citenamefont {Ye}, \citenamefont {Ketterle},\ and\ \citenamefont
  {Doyle}}]{hemmerling2016laser}%
  \BibitemOpen
  \bibfield  {author} {\bibinfo {author} {\bibfnamefont {B.}~\bibnamefont
  {Hemmerling}}, \bibinfo {author} {\bibfnamefont {E.}~\bibnamefont {Chae}},
  \bibinfo {author} {\bibfnamefont {A.}~\bibnamefont {Ravi}}, \bibinfo {author}
  {\bibfnamefont {L.}~\bibnamefont {Anderegg}}, \bibinfo {author}
  {\bibfnamefont {G.~K.}\ \bibnamefont {Drayna}}, \bibinfo {author}
  {\bibfnamefont {N.~R.}\ \bibnamefont {Hutzler}}, \bibinfo {author}
  {\bibfnamefont {A.~L.}\ \bibnamefont {Collopy}}, \bibinfo {author}
  {\bibfnamefont {J.}~\bibnamefont {Ye}}, \bibinfo {author} {\bibfnamefont
  {W.}~\bibnamefont {Ketterle}}, \ and\ \bibinfo {author} {\bibfnamefont
  {J.~M.}\ \bibnamefont {Doyle}},\ }\href@noop {} {\bibfield  {journal}
  {\bibinfo  {journal} {J. Phys. B}\ }\textbf {\bibinfo {volume} {49}},\
  \bibinfo {pages} {174001} (\bibinfo {year} {2016})}\BibitemShut {NoStop}%
\bibitem [{\citenamefont {Anderegg}\ \emph {et~al.}(2017)\citenamefont
  {Anderegg}, \citenamefont {Augenbraun}, \citenamefont {Chae}, \citenamefont
  {Hemmerling}, \citenamefont {Hutzler}, \citenamefont {Ravi}, \citenamefont
  {Collopy}, \citenamefont {Ye}, \citenamefont {Ketterle},\ and\ \citenamefont
  {Doyle}}]{anderegg2017radio}%
  \BibitemOpen
  \bibfield  {author} {\bibinfo {author} {\bibfnamefont {L.}~\bibnamefont
  {Anderegg}}, \bibinfo {author} {\bibfnamefont {B.~L.}\ \bibnamefont
  {Augenbraun}}, \bibinfo {author} {\bibfnamefont {E.}~\bibnamefont {Chae}},
  \bibinfo {author} {\bibfnamefont {B.}~\bibnamefont {Hemmerling}}, \bibinfo
  {author} {\bibfnamefont {N.~R.}\ \bibnamefont {Hutzler}}, \bibinfo {author}
  {\bibfnamefont {A.}~\bibnamefont {Ravi}}, \bibinfo {author} {\bibfnamefont
  {A.}~\bibnamefont {Collopy}}, \bibinfo {author} {\bibfnamefont
  {J.}~\bibnamefont {Ye}}, \bibinfo {author} {\bibfnamefont {W.}~\bibnamefont
  {Ketterle}}, \ and\ \bibinfo {author} {\bibfnamefont {J.~M.}\ \bibnamefont
  {Doyle}},\ }\href@noop {} {\bibfield  {journal} {\bibinfo  {journal} {Phys.
  Rev. Lett.}\ }\textbf {\bibinfo {volume} {119}},\ \bibinfo {pages} {103201}
  (\bibinfo {year} {2017})}\BibitemShut {NoStop}%
\bibitem [{\citenamefont {Lim}\ \emph {et~al.}(2018)\citenamefont {Lim},
  \citenamefont {Almond}, \citenamefont {Trigatzis}, \citenamefont {Devlin},
  \citenamefont {Fitch}, \citenamefont {Sauer}, \citenamefont {Tarbutt},\ and\
  \citenamefont {Hinds}}]{lim2018laser}%
  \BibitemOpen
  \bibfield  {author} {\bibinfo {author} {\bibfnamefont {J.}~\bibnamefont
  {Lim}}, \bibinfo {author} {\bibfnamefont {J.}~\bibnamefont {Almond}},
  \bibinfo {author} {\bibfnamefont {M.}~\bibnamefont {Trigatzis}}, \bibinfo
  {author} {\bibfnamefont {J.}~\bibnamefont {Devlin}}, \bibinfo {author}
  {\bibfnamefont {N.}~\bibnamefont {Fitch}}, \bibinfo {author} {\bibfnamefont
  {B.}~\bibnamefont {Sauer}}, \bibinfo {author} {\bibfnamefont
  {M.}~\bibnamefont {Tarbutt}}, \ and\ \bibinfo {author} {\bibfnamefont
  {E.}~\bibnamefont {Hinds}},\ }\href@noop {} {\bibfield  {journal} {\bibinfo
  {journal} {Phys. rev. lett.}\ }\textbf {\bibinfo {volume} {120}},\ \bibinfo
  {pages} {123201} (\bibinfo {year} {2018})}\BibitemShut {NoStop}%
\bibitem [{\citenamefont {McNally}\ \emph {et~al.}(2020)\citenamefont
  {McNally}, \citenamefont {Kozyryev}, \citenamefont {Vazquez-Carson},
  \citenamefont {Wenz}, \citenamefont {Wang},\ and\ \citenamefont
  {Zelevinsky}}]{mcnally2020optical}%
  \BibitemOpen
  \bibfield  {author} {\bibinfo {author} {\bibfnamefont {R.~L.}\ \bibnamefont
  {McNally}}, \bibinfo {author} {\bibfnamefont {I.}~\bibnamefont {Kozyryev}},
  \bibinfo {author} {\bibfnamefont {S.}~\bibnamefont {Vazquez-Carson}},
  \bibinfo {author} {\bibfnamefont {K.}~\bibnamefont {Wenz}}, \bibinfo {author}
  {\bibfnamefont {T.}~\bibnamefont {Wang}}, \ and\ \bibinfo {author}
  {\bibfnamefont {T.}~\bibnamefont {Zelevinsky}},\ }\href@noop {} {\bibfield
  {journal} {\bibinfo  {journal} {New J. Phys.}\ }\textbf {\bibinfo {volume}
  {22}},\ \bibinfo {pages} {083047} (\bibinfo {year} {2020})}\BibitemShut
  {NoStop}%
\bibitem [{\citenamefont {Baum}\ \emph {et~al.}(2020)\citenamefont {Baum},
  \citenamefont {Vilas}, \citenamefont {Hallas}, \citenamefont {Augenbraun},
  \citenamefont {Raval}, \citenamefont {Mitra},\ and\ \citenamefont
  {Doyle}}]{baum20201d}%
  \BibitemOpen
  \bibfield  {author} {\bibinfo {author} {\bibfnamefont {L.}~\bibnamefont
  {Baum}}, \bibinfo {author} {\bibfnamefont {N.~B.}\ \bibnamefont {Vilas}},
  \bibinfo {author} {\bibfnamefont {C.}~\bibnamefont {Hallas}}, \bibinfo
  {author} {\bibfnamefont {B.~L.}\ \bibnamefont {Augenbraun}}, \bibinfo
  {author} {\bibfnamefont {S.}~\bibnamefont {Raval}}, \bibinfo {author}
  {\bibfnamefont {D.}~\bibnamefont {Mitra}}, \ and\ \bibinfo {author}
  {\bibfnamefont {J.~M.}\ \bibnamefont {Doyle}},\ }\href@noop {} {\bibfield
  {journal} {\bibinfo  {journal} {Phys. Rev. Lett.}\ }\textbf {\bibinfo
  {volume} {124}},\ \bibinfo {pages} {133201} (\bibinfo {year}
  {2020})}\BibitemShut {NoStop}%
\bibitem [{\citenamefont {Mitra}\ \emph {et~al.}(2020)\citenamefont {Mitra},
  \citenamefont {Vilas}, \citenamefont {Hallas}, \citenamefont {Anderegg},
  \citenamefont {Augenbraun}, \citenamefont {Baum}, \citenamefont {Miller},
  \citenamefont {Raval},\ and\ \citenamefont {Doyle}}]{mitra2020direct}%
  \BibitemOpen
  \bibfield  {author} {\bibinfo {author} {\bibfnamefont {D.}~\bibnamefont
  {Mitra}}, \bibinfo {author} {\bibfnamefont {N.~B.}\ \bibnamefont {Vilas}},
  \bibinfo {author} {\bibfnamefont {C.}~\bibnamefont {Hallas}}, \bibinfo
  {author} {\bibfnamefont {L.}~\bibnamefont {Anderegg}}, \bibinfo {author}
  {\bibfnamefont {B.~L.}\ \bibnamefont {Augenbraun}}, \bibinfo {author}
  {\bibfnamefont {L.}~\bibnamefont {Baum}}, \bibinfo {author} {\bibfnamefont
  {C.}~\bibnamefont {Miller}}, \bibinfo {author} {\bibfnamefont
  {S.}~\bibnamefont {Raval}}, \ and\ \bibinfo {author} {\bibfnamefont {J.~M.}\
  \bibnamefont {Doyle}},\ }\href@noop {} {\bibfield  {journal} {\bibinfo
  {journal} {Science}\ }\textbf {\bibinfo {volume} {369}},\ \bibinfo {pages}
  {1366} (\bibinfo {year} {2020})}\BibitemShut {NoStop}%
\bibitem [{\citenamefont {Yeo}\ \emph {et~al.}(2015)\citenamefont {Yeo},
  \citenamefont {Hummon}, \citenamefont {Collopy}, \citenamefont {Yan},
  \citenamefont {Hemmerling}, \citenamefont {Chae}, \citenamefont {Doyle},\
  and\ \citenamefont {Ye}}]{yeo2015rotational}%
  \BibitemOpen
  \bibfield  {author} {\bibinfo {author} {\bibfnamefont {M.}~\bibnamefont
  {Yeo}}, \bibinfo {author} {\bibfnamefont {M.~T.}\ \bibnamefont {Hummon}},
  \bibinfo {author} {\bibfnamefont {A.~L.}\ \bibnamefont {Collopy}}, \bibinfo
  {author} {\bibfnamefont {B.}~\bibnamefont {Yan}}, \bibinfo {author}
  {\bibfnamefont {B.}~\bibnamefont {Hemmerling}}, \bibinfo {author}
  {\bibfnamefont {E.}~\bibnamefont {Chae}}, \bibinfo {author} {\bibfnamefont
  {J.~M.}\ \bibnamefont {Doyle}}, \ and\ \bibinfo {author} {\bibfnamefont
  {J.}~\bibnamefont {Ye}},\ }\href@noop {} {\bibfield  {journal} {\bibinfo
  {journal} {Phys. rev. lett.}\ }\textbf {\bibinfo {volume} {114}},\ \bibinfo
  {pages} {223003} (\bibinfo {year} {2015})}\BibitemShut {NoStop}%
\bibitem [{\citenamefont {Kozyryev}\ \emph {et~al.}(2017)\citenamefont
  {Kozyryev}, \citenamefont {Baum}, \citenamefont {Matsuda}, \citenamefont
  {Augenbraun}, \citenamefont {Anderegg}, \citenamefont {Sedlack},\ and\
  \citenamefont {Doyle}}]{kozyryev2017sisyphus}%
  \BibitemOpen
  \bibfield  {author} {\bibinfo {author} {\bibfnamefont {I.}~\bibnamefont
  {Kozyryev}}, \bibinfo {author} {\bibfnamefont {L.}~\bibnamefont {Baum}},
  \bibinfo {author} {\bibfnamefont {K.}~\bibnamefont {Matsuda}}, \bibinfo
  {author} {\bibfnamefont {B.~L.}\ \bibnamefont {Augenbraun}}, \bibinfo
  {author} {\bibfnamefont {L.}~\bibnamefont {Anderegg}}, \bibinfo {author}
  {\bibfnamefont {A.~P.}\ \bibnamefont {Sedlack}}, \ and\ \bibinfo {author}
  {\bibfnamefont {J.~M.}\ \bibnamefont {Doyle}},\ }\href@noop {} {\bibfield
  {journal} {\bibinfo  {journal} {Phys. Rev. Lett.}\ }\textbf {\bibinfo
  {volume} {118}},\ \bibinfo {pages} {173201} (\bibinfo {year}
  {2017})}\BibitemShut {NoStop}%
\bibitem [{\citenamefont {Augenbraun}\ \emph {et~al.}(2020)\citenamefont
  {Augenbraun}, \citenamefont {Lasner}, \citenamefont {Frenett}, \citenamefont
  {Sawaoka}, \citenamefont {Miller}, \citenamefont {Steimle},\ and\
  \citenamefont {Doyle}}]{augenbraun2020laser}%
  \BibitemOpen
  \bibfield  {author} {\bibinfo {author} {\bibfnamefont {B.~L.}\ \bibnamefont
  {Augenbraun}}, \bibinfo {author} {\bibfnamefont {Z.~D.}\ \bibnamefont
  {Lasner}}, \bibinfo {author} {\bibfnamefont {A.}~\bibnamefont {Frenett}},
  \bibinfo {author} {\bibfnamefont {H.}~\bibnamefont {Sawaoka}}, \bibinfo
  {author} {\bibfnamefont {C.}~\bibnamefont {Miller}}, \bibinfo {author}
  {\bibfnamefont {T.~C.}\ \bibnamefont {Steimle}}, \ and\ \bibinfo {author}
  {\bibfnamefont {J.~M.}\ \bibnamefont {Doyle}},\ }\href@noop {} {\bibfield
  {journal} {\bibinfo  {journal} {New J. Phys.}\ }\textbf {\bibinfo {volume}
  {22}},\ \bibinfo {pages} {022003} (\bibinfo {year} {2020})}\BibitemShut
  {NoStop}%
\bibitem [{\citenamefont {Novoselov}\ \emph {et~al.}(2004)\citenamefont
  {Novoselov}, \citenamefont {Geim}, \citenamefont {Morozov}, \citenamefont
  {Jiang}, \citenamefont {Zhang}, \citenamefont {Dubonos}, \citenamefont
  {Grigorieva},\ and\ \citenamefont {Firsov}}]{novoselov2004electric}%
  \BibitemOpen
  \bibfield  {author} {\bibinfo {author} {\bibfnamefont {K.~S.}\ \bibnamefont
  {Novoselov}}, \bibinfo {author} {\bibfnamefont {A.~K.}\ \bibnamefont {Geim}},
  \bibinfo {author} {\bibfnamefont {S.~V.}\ \bibnamefont {Morozov}}, \bibinfo
  {author} {\bibfnamefont {D.-e.}\ \bibnamefont {Jiang}}, \bibinfo {author}
  {\bibfnamefont {Y.}~\bibnamefont {Zhang}}, \bibinfo {author} {\bibfnamefont
  {S.~V.}\ \bibnamefont {Dubonos}}, \bibinfo {author} {\bibfnamefont {I.~V.}\
  \bibnamefont {Grigorieva}}, \ and\ \bibinfo {author} {\bibfnamefont {A.~A.}\
  \bibnamefont {Firsov}},\ }\href@noop {} {\bibfield  {journal} {\bibinfo
  {journal} {Science}\ }\textbf {\bibinfo {volume} {306}},\ \bibinfo {pages}
  {666} (\bibinfo {year} {2004})}\BibitemShut {NoStop}%
\bibitem [{\citenamefont {Geim}\ and\ \citenamefont
  {Novoselov}(2007)}]{geim2010rise}%
  \BibitemOpen
  \bibfield  {author} {\bibinfo {author} {\bibfnamefont {A.~K.}\ \bibnamefont
  {Geim}}\ and\ \bibinfo {author} {\bibfnamefont {K.~S.}\ \bibnamefont
  {Novoselov}},\ }\href@noop {} {\bibfield  {journal} {\bibinfo  {journal}
  {Nat. Mater.}\ }\textbf {\bibinfo {volume} {6}},\ \bibinfo {pages} {183–}
  (\bibinfo {year} {2007})}\BibitemShut {NoStop}%
\bibitem [{\citenamefont {O’Connell}(2018)}]{o2018carbon}%
  \BibitemOpen
  \bibfield  {author} {\bibinfo {author} {\bibfnamefont {M.~J.}\ \bibnamefont
  {O’Connell}},\ }\href@noop {} {\emph {\bibinfo {title} {Carbon nanotubes:
  properties and applications}}}\ (\bibinfo  {publisher} {CRC press},\ \bibinfo
  {year} {2018})\BibitemShut {NoStop}%
\bibitem [{\citenamefont {Dresselhaus}\ \emph {et~al.}(1996)\citenamefont
  {Dresselhaus}, \citenamefont {Dresselhaus},\ and\ \citenamefont
  {Eklund}}]{dresselhaus1996science}%
  \BibitemOpen
  \bibfield  {author} {\bibinfo {author} {\bibfnamefont {M.~S.}\ \bibnamefont
  {Dresselhaus}}, \bibinfo {author} {\bibfnamefont {G.}~\bibnamefont
  {Dresselhaus}}, \ and\ \bibinfo {author} {\bibfnamefont {P.~C.}\ \bibnamefont
  {Eklund}},\ }\href@noop {} {\emph {\bibinfo {title} {Science of fullerenes
  and carbon nanotubes}}}\ (\bibinfo  {publisher} {Elsevier},\ \bibinfo {year}
  {1996})\BibitemShut {NoStop}%
\bibitem [{\citenamefont {Changala}\ \emph {et~al.}(2019)\citenamefont
  {Changala}, \citenamefont {Weichman}, \citenamefont {Lee}, \citenamefont
  {Fermann},\ and\ \citenamefont {Ye}}]{changala2019rovibrational}%
  \BibitemOpen
  \bibfield  {author} {\bibinfo {author} {\bibfnamefont {P.~B.}\ \bibnamefont
  {Changala}}, \bibinfo {author} {\bibfnamefont {M.~L.}\ \bibnamefont
  {Weichman}}, \bibinfo {author} {\bibfnamefont {K.~F.}\ \bibnamefont {Lee}},
  \bibinfo {author} {\bibfnamefont {M.~E.}\ \bibnamefont {Fermann}}, \ and\
  \bibinfo {author} {\bibfnamefont {J.}~\bibnamefont {Ye}},\ }\href@noop {}
  {\bibfield  {journal} {\bibinfo  {journal} {Science}\ }\textbf {\bibinfo
  {volume} {363}},\ \bibinfo {pages} {49} (\bibinfo {year} {2019})}\BibitemShut
  {NoStop}%
\bibitem [{\citenamefont {McGuire}(2018)}]{mcguire20182018}%
  \BibitemOpen
  \bibfield  {author} {\bibinfo {author} {\bibfnamefont {B.~A.}\ \bibnamefont
  {McGuire}},\ }\href@noop {} {\bibfield  {journal} {\bibinfo  {journal}
  {Astrophys. J., Suppl. Ser.}\ }\textbf {\bibinfo {volume} {239}},\ \bibinfo
  {pages} {17} (\bibinfo {year} {2018})}\BibitemShut {NoStop}%
\bibitem [{\citenamefont {Ehrenfreund}\ and\ \citenamefont
  {Cami}(2010)}]{ehrenfreund2010cosmic}%
  \BibitemOpen
  \bibfield  {author} {\bibinfo {author} {\bibfnamefont {P.}~\bibnamefont
  {Ehrenfreund}}\ and\ \bibinfo {author} {\bibfnamefont {J.}~\bibnamefont
  {Cami}},\ }\href@noop {} {\bibfield  {journal} {\bibinfo  {journal} {Cold
  Spring Harb. Perspect. Biol.}\ }\textbf {\bibinfo {volume} {2}},\ \bibinfo
  {pages} {a002097} (\bibinfo {year} {2010})}\BibitemShut {NoStop}%
\bibitem [{\citenamefont {Ouellette}\ and\ \citenamefont
  {Rawn}(2018)}]{ouellette2018organic}%
  \BibitemOpen
  \bibfield  {author} {\bibinfo {author} {\bibfnamefont {R.~J.}\ \bibnamefont
  {Ouellette}}\ and\ \bibinfo {author} {\bibfnamefont {J.~D.}\ \bibnamefont
  {Rawn}},\ }\href@noop {} {\emph {\bibinfo {title} {Organic chemistry:
  structure, mechanism, synthesis}}}\ (\bibinfo  {publisher} {Academic Press},\
  \bibinfo {year} {2018})\BibitemShut {NoStop}%
\bibitem [{\citenamefont {Motaung}\ \emph {et~al.}(2010)\citenamefont
  {Motaung}, \citenamefont {Moodley}, \citenamefont {Manikandan},\ and\
  \citenamefont {Coville}}]{motaung2010situ}%
  \BibitemOpen
  \bibfield  {author} {\bibinfo {author} {\bibfnamefont {D.~E.}\ \bibnamefont
  {Motaung}}, \bibinfo {author} {\bibfnamefont {M.~K.}\ \bibnamefont
  {Moodley}}, \bibinfo {author} {\bibfnamefont {E.}~\bibnamefont {Manikandan}},
  \ and\ \bibinfo {author} {\bibfnamefont {N.~J.}\ \bibnamefont {Coville}},\
  }\href@noop {} {\bibfield  {journal} {\bibinfo  {journal} {J. Appl. Phys.}\
  }\textbf {\bibinfo {volume} {107}},\ \bibinfo {pages} {044308} (\bibinfo
  {year} {2010})}\BibitemShut {NoStop}%
\bibitem [{\citenamefont {Souza}\ and\ \citenamefont
  {LUTz}(1977)}]{souza1977detection}%
  \BibitemOpen
  \bibfield  {author} {\bibinfo {author} {\bibfnamefont {S.~P.}\ \bibnamefont
  {Souza}}\ and\ \bibinfo {author} {\bibfnamefont {B.~L.}\ \bibnamefont
  {LUTz}},\ }\href@noop {} {\bibfield  {journal} {\bibinfo  {journal} {The
  Astrophysical Journal}\ }\textbf {\bibinfo {volume} {216}},\ \bibinfo {pages}
  {L49} (\bibinfo {year} {1977})}\BibitemShut {NoStop}%
\bibitem [{\citenamefont {Kramida}\ \emph {et~al.}(2021)\citenamefont
  {Kramida}, \citenamefont {{Yu.~Ralchenko}}, \citenamefont {Reader},\ and\
  \citenamefont {{and NIST ASD Team}}}]{NIST_ASD}%
  \BibitemOpen
  \bibfield  {author} {\bibinfo {author} {\bibfnamefont {A.}~\bibnamefont
  {Kramida}}, \bibinfo {author} {\bibnamefont {{Yu.~Ralchenko}}}, \bibinfo
  {author} {\bibfnamefont {J.}~\bibnamefont {Reader}}, \ and\ \bibinfo {author}
  {\bibnamefont {{and NIST ASD Team}}},\ }\href@noop {} {}\bibinfo
  {howpublished} {{NIST Atomic Spectra Database (ver. 5.9), [Online].
  Available: {\tt{https://physics.nist.gov/asd}} [2021, November 9]. National
  Institute of Standards and Technology, Gaithersburg, MD.}} (\bibinfo {year}
  {2021})\BibitemShut {NoStop}%
\bibitem [{\citenamefont {Wells}\ and\ \citenamefont
  {Lane}(2011)}]{wells2011prospects}%
  \BibitemOpen
  \bibfield  {author} {\bibinfo {author} {\bibfnamefont {N.}~\bibnamefont
  {Wells}}\ and\ \bibinfo {author} {\bibfnamefont {I.~C.}\ \bibnamefont
  {Lane}},\ }\href@noop {} {\bibfield  {journal} {\bibinfo  {journal} {Phys.
  Chem. Chem. Phys.}\ }\textbf {\bibinfo {volume} {13}},\ \bibinfo {pages}
  {19036} (\bibinfo {year} {2011})}\BibitemShut {NoStop}%
\bibitem [{\citenamefont {Emsley}(1995)}]{emsley1995elements}%
  \BibitemOpen
  \bibfield  {author} {\bibinfo {author} {\bibfnamefont {J.}~\bibnamefont
  {Emsley}},\ }\href@noop {} {\emph {\bibinfo {title} {The Elements}}}\
  (\bibinfo  {publisher} {Oxford Chemistry Guides, Oxford Univ. Press},\
  \bibinfo {year} {1995})\BibitemShut {NoStop}%
\bibitem [{\citenamefont {Chen}\ \emph {et~al.}(2015)\citenamefont {Chen},
  \citenamefont {Kawaguchi}, \citenamefont {Bernath},\ and\ \citenamefont
  {Tang}}]{chen2015simultaneous}%
  \BibitemOpen
  \bibfield  {author} {\bibinfo {author} {\bibfnamefont {W.}~\bibnamefont
  {Chen}}, \bibinfo {author} {\bibfnamefont {K.}~\bibnamefont {Kawaguchi}},
  \bibinfo {author} {\bibfnamefont {P.~F.}\ \bibnamefont {Bernath}}, \ and\
  \bibinfo {author} {\bibfnamefont {J.}~\bibnamefont {Tang}},\ }\href@noop {}
  {\bibfield  {journal} {\bibinfo  {journal} {J. Chem. Phys.}\ }\textbf
  {\bibinfo {volume} {142}},\ \bibinfo {pages} {064317} (\bibinfo {year}
  {2015})}\BibitemShut {NoStop}%
\bibitem [{\citenamefont {Vogel}\ \emph {et~al.}(1999)\citenamefont {Vogel},
  \citenamefont {Dinneen}, \citenamefont {Gallagher},\ and\ \citenamefont
  {Hall}}]{vogel1999}%
  \BibitemOpen
  \bibfield  {author} {\bibinfo {author} {\bibfnamefont {K.~R.}\ \bibnamefont
  {Vogel}}, \bibinfo {author} {\bibfnamefont {T.~P.}\ \bibnamefont {Dinneen}},
  \bibinfo {author} {\bibfnamefont {A.}~\bibnamefont {Gallagher}}, \ and\
  \bibinfo {author} {\bibfnamefont {J.~L.}\ \bibnamefont {Hall}},\ }\href@noop
  {} {\bibfield  {journal} {\bibinfo  {journal} {IEEE Trans. Instrum. Meas.}\
  }\textbf {\bibinfo {volume} {48}},\ \bibinfo {pages} {618} (\bibinfo {year}
  {1999})}\BibitemShut {NoStop}%
\bibitem [{\citenamefont {Katori}\ \emph {et~al.}(1999)\citenamefont {Katori},
  \citenamefont {Ido}, \citenamefont {Isoya},\ and\ \citenamefont
  {Kuwata-Gonokami}}]{katori1999}%
  \BibitemOpen
  \bibfield  {author} {\bibinfo {author} {\bibfnamefont {H.}~\bibnamefont
  {Katori}}, \bibinfo {author} {\bibfnamefont {T.}~\bibnamefont {Ido}},
  \bibinfo {author} {\bibfnamefont {Y.}~\bibnamefont {Isoya}}, \ and\ \bibinfo
  {author} {\bibfnamefont {M.}~\bibnamefont {Kuwata-Gonokami}},\ }\href@noop {}
  {\bibfield  {journal} {\bibinfo  {journal} {Phys. Rev. Lett.}\ }\textbf
  {\bibinfo {volume} {82}},\ \bibinfo {pages} {1116} (\bibinfo {year}
  {1999})}\BibitemShut {NoStop}%
\bibitem [{\citenamefont {Kuwamoto}\ \emph {et~al.}(1999)\citenamefont
  {Kuwamoto}, \citenamefont {Honda}, \citenamefont {Takahashi},\ and\
  \citenamefont {Yabuzaki}}]{kuwamoto1999}%
  \BibitemOpen
  \bibfield  {author} {\bibinfo {author} {\bibfnamefont {T.}~\bibnamefont
  {Kuwamoto}}, \bibinfo {author} {\bibfnamefont {K.}~\bibnamefont {Honda}},
  \bibinfo {author} {\bibfnamefont {Y.}~\bibnamefont {Takahashi}}, \ and\
  \bibinfo {author} {\bibfnamefont {T.}~\bibnamefont {Yabuzaki}},\ }\href@noop
  {} {\bibfield  {journal} {\bibinfo  {journal} {Phys. Rev. A}\ }\textbf
  {\bibinfo {volume} {60}},\ \bibinfo {pages} {R745} (\bibinfo {year}
  {1999})}\BibitemShut {NoStop}%
\bibitem [{\citenamefont {Weltner~Jr}\ and\ \citenamefont
  {Van~Zee}(1989)}]{weltner1989carbon}%
  \BibitemOpen
  \bibfield  {author} {\bibinfo {author} {\bibfnamefont {W.}~\bibnamefont
  {Weltner~Jr}}\ and\ \bibinfo {author} {\bibfnamefont {R.~J.}\ \bibnamefont
  {Van~Zee}},\ }\href@noop {} {\bibfield  {journal} {\bibinfo  {journal} {Chem.
  Rev.}\ }\textbf {\bibinfo {volume} {89}},\ \bibinfo {pages} {1713} (\bibinfo
  {year} {1989})}\BibitemShut {NoStop}%
\bibitem [{\citenamefont {Brooke}\ \emph {et~al.}(2013)\citenamefont {Brooke},
  \citenamefont {Bernath}, \citenamefont {Schmidt},\ and\ \citenamefont
  {Bacskay}}]{brooke2013line}%
  \BibitemOpen
  \bibfield  {author} {\bibinfo {author} {\bibfnamefont {J.~S.}\ \bibnamefont
  {Brooke}}, \bibinfo {author} {\bibfnamefont {P.~F.}\ \bibnamefont {Bernath}},
  \bibinfo {author} {\bibfnamefont {T.~W.}\ \bibnamefont {Schmidt}}, \ and\
  \bibinfo {author} {\bibfnamefont {G.~B.}\ \bibnamefont {Bacskay}},\
  }\href@noop {} {\bibfield  {journal} {\bibinfo  {journal} {J. Quant.
  Spectrosc. Radiat. Transfer}\ }\textbf {\bibinfo {volume} {124}},\ \bibinfo
  {pages} {11} (\bibinfo {year} {2013})}\BibitemShut {NoStop}%
\bibitem [{\citenamefont {Kokkin}\ \emph {et~al.}(2007)\citenamefont {Kokkin},
  \citenamefont {Bacskay},\ and\ \citenamefont
  {Schmidt}}]{kokkin2007oscillator}%
  \BibitemOpen
  \bibfield  {author} {\bibinfo {author} {\bibfnamefont {D.~L.}\ \bibnamefont
  {Kokkin}}, \bibinfo {author} {\bibfnamefont {G.~B.}\ \bibnamefont {Bacskay}},
  \ and\ \bibinfo {author} {\bibfnamefont {T.~W.}\ \bibnamefont {Schmidt}},\
  }\href@noop {} {\bibfield  {journal} {\bibinfo  {journal} {J. Chem. Phys.}\
  }\textbf {\bibinfo {volume} {126}},\ \bibinfo {pages} {084302} (\bibinfo
  {year} {2007})}\BibitemShut {NoStop}%
\bibitem [{\citenamefont {Furtenbacher}\ \emph {et~al.}(2016)\citenamefont
  {Furtenbacher}, \citenamefont {Szab{\'o}}, \citenamefont {Cs{\'a}sz{\'a}r},
  \citenamefont {Bernath}, \citenamefont {Yurchenko},\ and\ \citenamefont
  {Tennyson}}]{furtenbacher2016experimental}%
  \BibitemOpen
  \bibfield  {author} {\bibinfo {author} {\bibfnamefont {T.}~\bibnamefont
  {Furtenbacher}}, \bibinfo {author} {\bibfnamefont {I.}~\bibnamefont
  {Szab{\'o}}}, \bibinfo {author} {\bibfnamefont {A.~G.}\ \bibnamefont
  {Cs{\'a}sz{\'a}r}}, \bibinfo {author} {\bibfnamefont {P.~F.}\ \bibnamefont
  {Bernath}}, \bibinfo {author} {\bibfnamefont {S.~N.}\ \bibnamefont
  {Yurchenko}}, \ and\ \bibinfo {author} {\bibfnamefont {J.}~\bibnamefont
  {Tennyson}},\ }\href@noop {} {\bibfield  {journal} {\bibinfo  {journal}
  {Astrophys. J. Suppl. Ser.}\ }\textbf {\bibinfo {volume} {224}},\ \bibinfo
  {pages} {44} (\bibinfo {year} {2016})}\BibitemShut {NoStop}%
\bibitem [{\citenamefont {Abrams}\ and\ \citenamefont
  {Sherrill}(2004)}]{abrams2004full}%
  \BibitemOpen
  \bibfield  {author} {\bibinfo {author} {\bibfnamefont {M.~L.}\ \bibnamefont
  {Abrams}}\ and\ \bibinfo {author} {\bibfnamefont {C.~D.}\ \bibnamefont
  {Sherrill}},\ }\href@noop {} {\bibfield  {journal} {\bibinfo  {journal} {J.
  Chem. Phys.}\ }\textbf {\bibinfo {volume} {121}},\ \bibinfo {pages} {9211}
  (\bibinfo {year} {2004})}\BibitemShut {NoStop}%
\bibitem [{\citenamefont {Tennyson}\ \emph {et~al.}(2016)\citenamefont
  {Tennyson}, \citenamefont {Yurchenko}, \citenamefont {Al-Refaie},
  \citenamefont {Barton}, \citenamefont {Chubb}, \citenamefont {Coles},
  \citenamefont {Diamantopoulou}, \citenamefont {Gorman}, \citenamefont {Hill},
  \citenamefont {Lam} \emph {et~al.}}]{tennyson2016exomol}%
  \BibitemOpen
  \bibfield  {author} {\bibinfo {author} {\bibfnamefont {J.}~\bibnamefont
  {Tennyson}}, \bibinfo {author} {\bibfnamefont {S.~N.}\ \bibnamefont
  {Yurchenko}}, \bibinfo {author} {\bibfnamefont {A.~F.}\ \bibnamefont
  {Al-Refaie}}, \bibinfo {author} {\bibfnamefont {E.~J.}\ \bibnamefont
  {Barton}}, \bibinfo {author} {\bibfnamefont {K.~L.}\ \bibnamefont {Chubb}},
  \bibinfo {author} {\bibfnamefont {P.~A.}\ \bibnamefont {Coles}}, \bibinfo
  {author} {\bibfnamefont {S.}~\bibnamefont {Diamantopoulou}}, \bibinfo
  {author} {\bibfnamefont {M.~N.}\ \bibnamefont {Gorman}}, \bibinfo {author}
  {\bibfnamefont {C.}~\bibnamefont {Hill}}, \bibinfo {author} {\bibfnamefont
  {A.~Z.}\ \bibnamefont {Lam}},  \emph {et~al.},\ }\href@noop {} {\bibfield
  {journal} {\bibinfo  {journal} {J. Mol. Spectrosc.}\ }\textbf {\bibinfo
  {volume} {327}},\ \bibinfo {pages} {73} (\bibinfo {year} {2016})}\BibitemShut
  {NoStop}%
\bibitem [{\citenamefont {Yurchenko}\ \emph {et~al.}(2018)\citenamefont
  {Yurchenko}, \citenamefont {Szab{\'o}}, \citenamefont {Pyatenko},\ and\
  \citenamefont {Tennyson}}]{yurchenko2018exomol}%
  \BibitemOpen
  \bibfield  {author} {\bibinfo {author} {\bibfnamefont {S.~N.}\ \bibnamefont
  {Yurchenko}}, \bibinfo {author} {\bibfnamefont {I.}~\bibnamefont
  {Szab{\'o}}}, \bibinfo {author} {\bibfnamefont {E.}~\bibnamefont {Pyatenko}},
  \ and\ \bibinfo {author} {\bibfnamefont {J.}~\bibnamefont {Tennyson}},\
  }\href@noop {} {\bibfield  {journal} {\bibinfo  {journal} {Mon. Not. R.
  Astron. Soc.}\ }\textbf {\bibinfo {volume} {480}},\ \bibinfo {pages} {3397}
  (\bibinfo {year} {2018})}\BibitemShut {NoStop}%
\bibitem [{\citenamefont {McKemmish}\ \emph {et~al.}(2020)\citenamefont
  {McKemmish}, \citenamefont {Syme}, \citenamefont {Borsovszky}, \citenamefont
  {Yurchenko}, \citenamefont {Tennyson}, \citenamefont {Furtenbacher},\ and\
  \citenamefont {Cs{\'a}sz{\'a}r}}]{mckemmish2020update}%
  \BibitemOpen
  \bibfield  {author} {\bibinfo {author} {\bibfnamefont {L.~K.}\ \bibnamefont
  {McKemmish}}, \bibinfo {author} {\bibfnamefont {A.-M.}\ \bibnamefont {Syme}},
  \bibinfo {author} {\bibfnamefont {J.}~\bibnamefont {Borsovszky}}, \bibinfo
  {author} {\bibfnamefont {S.~N.}\ \bibnamefont {Yurchenko}}, \bibinfo {author}
  {\bibfnamefont {J.}~\bibnamefont {Tennyson}}, \bibinfo {author}
  {\bibfnamefont {T.}~\bibnamefont {Furtenbacher}}, \ and\ \bibinfo {author}
  {\bibfnamefont {A.~G.}\ \bibnamefont {Cs{\'a}sz{\'a}r}},\ }\href@noop {}
  {\bibfield  {journal} {\bibinfo  {journal} {Mon. Notices Royal Astron. Soc.}\
  }\textbf {\bibinfo {volume} {497}},\ \bibinfo {pages} {1081} (\bibinfo {year}
  {2020})}\BibitemShut {NoStop}%
\bibitem [{Exo()}]{ExoMol}%
  \BibitemOpen
  \href {https://www.exomol.com/data/licence/} {\enquote {\bibinfo {title}
  {Exomol, https://www.exomol.com/},}\ }\BibitemShut {NoStop}%
\bibitem [{SI()}]{SI}%
  \BibitemOpen
  \href@noop {} {}\bibinfo {note} {See Supplemental Material for details on the
  calculation of branching ratios, cycling closure, and deflection and cooling
  parameters.}\BibitemShut {Stop}%
\bibitem [{\citenamefont {Lefebvre-Brion}\ and\ \citenamefont
  {Field}(2004)}]{lefebvre2004spectra}%
  \BibitemOpen
  \bibfield  {author} {\bibinfo {author} {\bibfnamefont {H.}~\bibnamefont
  {Lefebvre-Brion}}\ and\ \bibinfo {author} {\bibfnamefont {R.~W.}\
  \bibnamefont {Field}},\ }\href@noop {} {\emph {\bibinfo {title} {The Spectra
  and Dynamics of Diatomic Molecules}}}\ (\bibinfo  {publisher} {Elsevier},\
  \bibinfo {year} {2004})\BibitemShut {NoStop}%
\bibitem [{\citenamefont {Brown}\ \emph {et~al.}(2003)\citenamefont {Brown},
  \citenamefont {Brown},\ and\ \citenamefont
  {Carrington}}]{brown2003rotational}%
  \BibitemOpen
  \bibfield  {author} {\bibinfo {author} {\bibfnamefont {J.~M.}\ \bibnamefont
  {Brown}}, \bibinfo {author} {\bibfnamefont {J.~M.}\ \bibnamefont {Brown}}, \
  and\ \bibinfo {author} {\bibfnamefont {A.}~\bibnamefont {Carrington}},\
  }\href@noop {} {\emph {\bibinfo {title} {Rotational spectroscopy of diatomic
  molecules}}}\ (\bibinfo  {publisher} {Cambridge university press},\ \bibinfo
  {year} {2003})\BibitemShut {NoStop}%
\bibitem [{\citenamefont {Di~Rosa}(2004)}]{di2004laser}%
  \BibitemOpen
  \bibfield  {author} {\bibinfo {author} {\bibfnamefont {M.}~\bibnamefont
  {Di~Rosa}},\ }\href@noop {} {\bibfield  {journal} {\bibinfo  {journal} {Eur.
  Phys. J. D}\ }\textbf {\bibinfo {volume} {31}},\ \bibinfo {pages} {395}
  (\bibinfo {year} {2004})}\BibitemShut {NoStop}%
\bibitem [{\citenamefont {Tarbutt}\ \emph {et~al.}(2013)\citenamefont
  {Tarbutt}, \citenamefont {Sauer}, \citenamefont {Hudson},\ and\ \citenamefont
  {Hinds}}]{tarbutt2013design}%
  \BibitemOpen
  \bibfield  {author} {\bibinfo {author} {\bibfnamefont {M.}~\bibnamefont
  {Tarbutt}}, \bibinfo {author} {\bibfnamefont {B.}~\bibnamefont {Sauer}},
  \bibinfo {author} {\bibfnamefont {J.}~\bibnamefont {Hudson}}, \ and\ \bibinfo
  {author} {\bibfnamefont {E.}~\bibnamefont {Hinds}},\ }\href@noop {}
  {\bibfield  {journal} {\bibinfo  {journal} {New J. Phys.}\ }\textbf {\bibinfo
  {volume} {15}},\ \bibinfo {pages} {053034} (\bibinfo {year}
  {2013})}\BibitemShut {NoStop}%
\bibitem [{\citenamefont {Maxwell}\ \emph {et~al.}(2005)\citenamefont
  {Maxwell}, \citenamefont {Brahms}, \citenamefont {deCarvalho}, \citenamefont
  {Glenn}, \citenamefont {Helton}, \citenamefont {Nguyen}, \citenamefont
  {Patterson}, \citenamefont {Petricka}, \citenamefont {DeMille},\ and\
  \citenamefont {Doyle}}]{maxwell2005high}%
  \BibitemOpen
  \bibfield  {author} {\bibinfo {author} {\bibfnamefont {S.~E.}\ \bibnamefont
  {Maxwell}}, \bibinfo {author} {\bibfnamefont {N.}~\bibnamefont {Brahms}},
  \bibinfo {author} {\bibfnamefont {R.}~\bibnamefont {deCarvalho}}, \bibinfo
  {author} {\bibfnamefont {D.~R.}\ \bibnamefont {Glenn}}, \bibinfo {author}
  {\bibfnamefont {J.}~\bibnamefont {Helton}}, \bibinfo {author} {\bibfnamefont
  {S.~V.}\ \bibnamefont {Nguyen}}, \bibinfo {author} {\bibfnamefont
  {D.}~\bibnamefont {Patterson}}, \bibinfo {author} {\bibfnamefont
  {J.}~\bibnamefont {Petricka}}, \bibinfo {author} {\bibfnamefont
  {D.}~\bibnamefont {DeMille}}, \ and\ \bibinfo {author} {\bibfnamefont
  {J.~M.}\ \bibnamefont {Doyle}},\ }\href@noop {} {\bibfield  {journal}
  {\bibinfo  {journal} {Phys. Rev. Lett.}\ }\textbf {\bibinfo {volume} {95}},\
  \bibinfo {pages} {173201} (\bibinfo {year} {2005})}\BibitemShut {NoStop}%
\bibitem [{\citenamefont {Hutzler}\ \emph {et~al.}(2012)\citenamefont
  {Hutzler}, \citenamefont {Lu},\ and\ \citenamefont
  {Doyle}}]{hutzler2012buffer}%
  \BibitemOpen
  \bibfield  {author} {\bibinfo {author} {\bibfnamefont {N.~R.}\ \bibnamefont
  {Hutzler}}, \bibinfo {author} {\bibfnamefont {H.-I.}\ \bibnamefont {Lu}}, \
  and\ \bibinfo {author} {\bibfnamefont {J.~M.}\ \bibnamefont {Doyle}},\
  }\href@noop {} {\bibfield  {journal} {\bibinfo  {journal} {Chem. Rev.}\
  }\textbf {\bibinfo {volume} {112}},\ \bibinfo {pages} {4803} (\bibinfo {year}
  {2012})}\BibitemShut {NoStop}%
\bibitem [{\citenamefont {Phillips}\ and\ \citenamefont
  {Metcalf}(1982)}]{phillips1982laser}%
  \BibitemOpen
  \bibfield  {author} {\bibinfo {author} {\bibfnamefont {W.~D.}\ \bibnamefont
  {Phillips}}\ and\ \bibinfo {author} {\bibfnamefont {H.}~\bibnamefont
  {Metcalf}},\ }\href@noop {} {\bibfield  {journal} {\bibinfo  {journal} {Phys.
  Rev. Lett.}\ }\textbf {\bibinfo {volume} {48}},\ \bibinfo {pages} {596}
  (\bibinfo {year} {1982})}\BibitemShut {NoStop}%
\bibitem [{\citenamefont {Campbell}\ and\ \citenamefont
  {Doyle}(2009)}]{campbell2009cooling}%
  \BibitemOpen
  \bibfield  {author} {\bibinfo {author} {\bibfnamefont {W.~C.}\ \bibnamefont
  {Campbell}}\ and\ \bibinfo {author} {\bibfnamefont {J.~M.}\ \bibnamefont
  {Doyle}},\ }in\ \href@noop {} {\emph {\bibinfo {booktitle} {Cold
  Molecules}}}\ (\bibinfo  {publisher} {CRC Press},\ \bibinfo {year} {2009})\
  pp.\ \bibinfo {pages} {473--508}\BibitemShut {NoStop}%
\bibitem [{\citenamefont {Kaiser}\ and\ \citenamefont
  {Suits}(1995)}]{kaiser1995high}%
  \BibitemOpen
  \bibfield  {author} {\bibinfo {author} {\bibfnamefont {R.}~\bibnamefont
  {Kaiser}}\ and\ \bibinfo {author} {\bibfnamefont {A.~G.}\ \bibnamefont
  {Suits}},\ }\href@noop {} {\bibfield  {journal} {\bibinfo  {journal} {Rev.
  Sci. Instrum.}\ }\textbf {\bibinfo {volume} {66}},\ \bibinfo {pages} {5405}
  (\bibinfo {year} {1995})}\BibitemShut {NoStop}%
\bibitem [{\citenamefont {Ursu}\ \emph {et~al.}(2018)\citenamefont {Ursu},
  \citenamefont {Nica},\ and\ \citenamefont {Focsa}}]{ursu2018excimer}%
  \BibitemOpen
  \bibfield  {author} {\bibinfo {author} {\bibfnamefont {C.}~\bibnamefont
  {Ursu}}, \bibinfo {author} {\bibfnamefont {P.}~\bibnamefont {Nica}}, \ and\
  \bibinfo {author} {\bibfnamefont {C.}~\bibnamefont {Focsa}},\ }\href@noop {}
  {\bibfield  {journal} {\bibinfo  {journal} {Appl. Surf. Sci.}\ }\textbf
  {\bibinfo {volume} {456}},\ \bibinfo {pages} {717} (\bibinfo {year}
  {2018})}\BibitemShut {NoStop}%
\bibitem [{\citenamefont {Savi{\'c}}\ \emph {et~al.}(2005)\citenamefont
  {Savi{\'c}}, \citenamefont {{\v{C}}erm{\'a}k},\ and\ \citenamefont
  {Gerlich}}]{savic2005reactions}%
  \BibitemOpen
  \bibfield  {author} {\bibinfo {author} {\bibfnamefont {I.}~\bibnamefont
  {Savi{\'c}}}, \bibinfo {author} {\bibfnamefont {I.}~\bibnamefont
  {{\v{C}}erm{\'a}k}}, \ and\ \bibinfo {author} {\bibfnamefont
  {D.}~\bibnamefont {Gerlich}},\ }\href@noop {} {\bibfield  {journal} {\bibinfo
   {journal} {Inte. J. Mass Spectrom.}\ }\textbf {\bibinfo {volume} {240}},\
  \bibinfo {pages} {139} (\bibinfo {year} {2005})}\BibitemShut {NoStop}%
\bibitem [{\citenamefont {Mulliken}(1939)}]{mulliken1939note}%
  \BibitemOpen
  \bibfield  {author} {\bibinfo {author} {\bibfnamefont {R.~S.}\ \bibnamefont
  {Mulliken}},\ }\href@noop {} {\bibfield  {journal} {\bibinfo  {journal}
  {Phys. rev.}\ }\textbf {\bibinfo {volume} {56}},\ \bibinfo {pages} {778}
  (\bibinfo {year} {1939})}\BibitemShut {NoStop}%
\bibitem [{\citenamefont {Krause}(1979)}]{krause1979carbon}%
  \BibitemOpen
  \bibfield  {author} {\bibinfo {author} {\bibfnamefont {H.}~\bibnamefont
  {Krause}},\ }\href@noop {} {\bibfield  {journal} {\bibinfo  {journal} {J.
  Chem. Phys.}\ }\textbf {\bibinfo {volume} {70}},\ \bibinfo {pages} {3871}
  (\bibinfo {year} {1979})}\BibitemShut {NoStop}%
\bibitem [{\citenamefont {Najar}\ \emph {et~al.}(2008)\citenamefont {Najar},
  \citenamefont {Abdallah}, \citenamefont {Jaidane},\ and\ \citenamefont
  {Lakhdar}}]{najar2008potential}%
  \BibitemOpen
  \bibfield  {author} {\bibinfo {author} {\bibfnamefont {F.}~\bibnamefont
  {Najar}}, \bibinfo {author} {\bibfnamefont {D.~B.}\ \bibnamefont {Abdallah}},
  \bibinfo {author} {\bibfnamefont {N.}~\bibnamefont {Jaidane}}, \ and\
  \bibinfo {author} {\bibfnamefont {Z.~B.}\ \bibnamefont {Lakhdar}},\
  }\href@noop {} {\bibfield  {journal} {\bibinfo  {journal} {Chem. Phys.
  Lett.}\ }\textbf {\bibinfo {volume} {460}},\ \bibinfo {pages} {31} (\bibinfo
  {year} {2008})}\BibitemShut {NoStop}%
\bibitem [{\citenamefont {P{\'a}ramo}\ \emph {et~al.}(2006)\citenamefont
  {P{\'a}ramo}, \citenamefont {Canosa}, \citenamefont {Le~Picard},\ and\
  \citenamefont {Sims}}]{paramo2006experimental}%
  \BibitemOpen
  \bibfield  {author} {\bibinfo {author} {\bibfnamefont {A.}~\bibnamefont
  {P{\'a}ramo}}, \bibinfo {author} {\bibfnamefont {A.}~\bibnamefont {Canosa}},
  \bibinfo {author} {\bibfnamefont {S.~D.}\ \bibnamefont {Le~Picard}}, \ and\
  \bibinfo {author} {\bibfnamefont {I.~R.}\ \bibnamefont {Sims}},\ }\href@noop
  {} {\bibfield  {journal} {\bibinfo  {journal} {J. Phys. Chem. A}\ }\textbf
  {\bibinfo {volume} {110}},\ \bibinfo {pages} {3121} (\bibinfo {year}
  {2006})}\BibitemShut {NoStop}%
\bibitem [{\citenamefont {Jadbabaie}\ \emph {et~al.}(2020)\citenamefont
  {Jadbabaie}, \citenamefont {Pilgram}, \citenamefont {K{\l}os}, \citenamefont
  {Kotochigova},\ and\ \citenamefont {Hutzler}}]{jadbabaie2020enhanced}%
  \BibitemOpen
  \bibfield  {author} {\bibinfo {author} {\bibfnamefont {A.}~\bibnamefont
  {Jadbabaie}}, \bibinfo {author} {\bibfnamefont {N.~H.}\ \bibnamefont
  {Pilgram}}, \bibinfo {author} {\bibfnamefont {J.}~\bibnamefont {K{\l}os}},
  \bibinfo {author} {\bibfnamefont {S.}~\bibnamefont {Kotochigova}}, \ and\
  \bibinfo {author} {\bibfnamefont {N.~R.}\ \bibnamefont {Hutzler}},\
  }\href@noop {} {\bibfield  {journal} {\bibinfo  {journal} {New J. Phys.}\
  }\textbf {\bibinfo {volume} {22}},\ \bibinfo {pages} {022002} (\bibinfo
  {year} {2020})}\BibitemShut {NoStop}%
\bibitem [{\citenamefont {Anderegg}\ \emph {et~al.}(2019)\citenamefont
  {Anderegg}, \citenamefont {Cheuk}, \citenamefont {Bao}, \citenamefont
  {Burchesky}, \citenamefont {Ketterle}, \citenamefont {Ni},\ and\
  \citenamefont {Doyle}}]{anderegg2019optical}%
  \BibitemOpen
  \bibfield  {author} {\bibinfo {author} {\bibfnamefont {L.}~\bibnamefont
  {Anderegg}}, \bibinfo {author} {\bibfnamefont {L.~W.}\ \bibnamefont {Cheuk}},
  \bibinfo {author} {\bibfnamefont {Y.}~\bibnamefont {Bao}}, \bibinfo {author}
  {\bibfnamefont {S.}~\bibnamefont {Burchesky}}, \bibinfo {author}
  {\bibfnamefont {W.}~\bibnamefont {Ketterle}}, \bibinfo {author}
  {\bibfnamefont {K.-K.}\ \bibnamefont {Ni}}, \ and\ \bibinfo {author}
  {\bibfnamefont {J.~M.}\ \bibnamefont {Doyle}},\ }\href@noop {} {\bibfield
  {journal} {\bibinfo  {journal} {Science}\ }\textbf {\bibinfo {volume}
  {365}},\ \bibinfo {pages} {1156} (\bibinfo {year} {2019})}\BibitemShut
  {NoStop}%
\bibitem [{\citenamefont {Tarbutt}(2015)}]{tarbutt2015magneto}%
  \BibitemOpen
  \bibfield  {author} {\bibinfo {author} {\bibfnamefont {M.}~\bibnamefont
  {Tarbutt}},\ }\href@noop {} {\bibfield  {journal} {\bibinfo  {journal} {New
  J. Phys.}\ }\textbf {\bibinfo {volume} {17}},\ \bibinfo {pages} {015007}
  (\bibinfo {year} {2015})}\BibitemShut {NoStop}%
\bibitem [{\citenamefont {S{\"o}ding}\ \emph {et~al.}(1997)\citenamefont
  {S{\"o}ding}, \citenamefont {Grimm}, \citenamefont {Ovchinnikov},
  \citenamefont {Bouyer},\ and\ \citenamefont {Salomon}}]{soding1997short}%
  \BibitemOpen
  \bibfield  {author} {\bibinfo {author} {\bibfnamefont {J.}~\bibnamefont
  {S{\"o}ding}}, \bibinfo {author} {\bibfnamefont {R.}~\bibnamefont {Grimm}},
  \bibinfo {author} {\bibfnamefont {Y.~B.}\ \bibnamefont {Ovchinnikov}},
  \bibinfo {author} {\bibfnamefont {P.}~\bibnamefont {Bouyer}}, \ and\ \bibinfo
  {author} {\bibfnamefont {C.}~\bibnamefont {Salomon}},\ }\href@noop {}
  {\bibfield  {journal} {\bibinfo  {journal} {Phys. rev. lett.}\ }\textbf
  {\bibinfo {volume} {78}},\ \bibinfo {pages} {1420} (\bibinfo {year}
  {1997})}\BibitemShut {NoStop}%
\bibitem [{\citenamefont {Chieda}\ and\ \citenamefont
  {Eyler}(2011)}]{chieda2011prospects}%
  \BibitemOpen
  \bibfield  {author} {\bibinfo {author} {\bibfnamefont {M.}~\bibnamefont
  {Chieda}}\ and\ \bibinfo {author} {\bibfnamefont {E.}~\bibnamefont {Eyler}},\
  }\href@noop {} {\bibfield  {journal} {\bibinfo  {journal} {Phys. Rev. A}\
  }\textbf {\bibinfo {volume} {84}},\ \bibinfo {pages} {063401} (\bibinfo
  {year} {2011})}\BibitemShut {NoStop}%
\bibitem [{\citenamefont {Kozyryev}\ \emph {et~al.}(2018)\citenamefont
  {Kozyryev}, \citenamefont {Baum}, \citenamefont {Aldridge}, \citenamefont
  {Yu}, \citenamefont {Eyler},\ and\ \citenamefont
  {Doyle}}]{kozyryev2018coherent}%
  \BibitemOpen
  \bibfield  {author} {\bibinfo {author} {\bibfnamefont {I.}~\bibnamefont
  {Kozyryev}}, \bibinfo {author} {\bibfnamefont {L.}~\bibnamefont {Baum}},
  \bibinfo {author} {\bibfnamefont {L.}~\bibnamefont {Aldridge}}, \bibinfo
  {author} {\bibfnamefont {P.}~\bibnamefont {Yu}}, \bibinfo {author}
  {\bibfnamefont {E.~E.}\ \bibnamefont {Eyler}}, \ and\ \bibinfo {author}
  {\bibfnamefont {J.~M.}\ \bibnamefont {Doyle}},\ }\href@noop {} {\bibfield
  {journal} {\bibinfo  {journal} {Phys. Rev. Rett.}\ }\textbf {\bibinfo
  {volume} {120}},\ \bibinfo {pages} {063205} (\bibinfo {year}
  {2018})}\BibitemShut {NoStop}%
\bibitem [{\citenamefont {Wenz}\ \emph {et~al.}(2020)\citenamefont {Wenz},
  \citenamefont {Kozyryev}, \citenamefont {McNally}, \citenamefont {Aldridge},\
  and\ \citenamefont {Zelevinsky}}]{wenz2020large}%
  \BibitemOpen
  \bibfield  {author} {\bibinfo {author} {\bibfnamefont {K.}~\bibnamefont
  {Wenz}}, \bibinfo {author} {\bibfnamefont {I.}~\bibnamefont {Kozyryev}},
  \bibinfo {author} {\bibfnamefont {R.~L.}\ \bibnamefont {McNally}}, \bibinfo
  {author} {\bibfnamefont {L.}~\bibnamefont {Aldridge}}, \ and\ \bibinfo
  {author} {\bibfnamefont {T.}~\bibnamefont {Zelevinsky}},\ }\href@noop {}
  {\bibfield  {journal} {\bibinfo  {journal} {Phys. Rev. Res.}\ }\textbf
  {\bibinfo {volume} {2}},\ \bibinfo {pages} {043377} (\bibinfo {year}
  {2020})}\BibitemShut {NoStop}%
\end{thebibliography}%


\begin{thebibliography}{7}%
\makeatletter
\providecommand \@ifxundefined [1]{%
 \@ifx{#1\undefined}
}%
\providecommand \@ifnum [1]{%
 \ifnum #1\expandafter \@firstoftwo
 \else \expandafter \@secondoftwo
 \fi
}%
\providecommand \@ifx [1]{%
 \ifx #1\expandafter \@firstoftwo
 \else \expandafter \@secondoftwo
 \fi
}%
\providecommand \natexlab [1]{#1}%
\providecommand \enquote  [1]{``#1''}%
\providecommand \bibnamefont  [1]{#1}%
\providecommand \bibfnamefont [1]{#1}%
\providecommand \citenamefont [1]{#1}%
\providecommand \href@noop [0]{\@secondoftwo}%
\providecommand \href [0]{\begingroup \@sanitize@url \@href}%
\providecommand \@href[1]{\@@startlink{#1}\@@href}%
\providecommand \@@href[1]{\endgroup#1\@@endlink}%
\providecommand \@sanitize@url [0]{\catcode `\\12\catcode `\$12\catcode
  `\&12\catcode `\#12\catcode `\^12\catcode `\_12\catcode `\%12\relax}%
\providecommand \@@startlink[1]{}%
\providecommand \@@endlink[0]{}%
\providecommand \url  [0]{\begingroup\@sanitize@url \@url }%
\providecommand \@url [1]{\endgroup\@href {#1}{\urlprefix }}%
\providecommand \urlprefix  [0]{URL }%
\providecommand \Eprint [0]{\href }%
\providecommand \doibase [0]{http://dx.doi.org/}%
\providecommand \selectlanguage [0]{\@gobble}%
\providecommand \bibinfo  [0]{\@secondoftwo}%
\providecommand \bibfield  [0]{\@secondoftwo}%
\providecommand \translation [1]{[#1]}%
\providecommand \BibitemOpen [0]{}%
\providecommand \bibitemStop [0]{}%
\providecommand \bibitemNoStop [0]{.\EOS\space}%
\providecommand \EOS [0]{\spacefactor3000\relax}%
\providecommand \BibitemShut  [1]{\csname bibitem#1\endcsname}%
\let\auto@bib@innerbib\@empty
\bibitem [{\citenamefont {Tennyson}\ \emph {et~al.}(2016)\citenamefont
  {Tennyson}, \citenamefont {Yurchenko}, \citenamefont {Al-Refaie},
  \citenamefont {Barton}, \citenamefont {Chubb}, \citenamefont {Coles},
  \citenamefont {Diamantopoulou}, \citenamefont {Gorman}, \citenamefont {Hill},
  \citenamefont {Lam} \emph {et~al.}}]{tennyson2016exomol}%
  \BibitemOpen
  \bibfield  {author} {\bibinfo {author} {\bibfnamefont {J.}~\bibnamefont
  {Tennyson}}, \bibinfo {author} {\bibfnamefont {S.~N.}\ \bibnamefont
  {Yurchenko}}, \bibinfo {author} {\bibfnamefont {A.~F.}\ \bibnamefont
  {Al-Refaie}}, \bibinfo {author} {\bibfnamefont {E.~J.}\ \bibnamefont
  {Barton}}, \bibinfo {author} {\bibfnamefont {K.~L.}\ \bibnamefont {Chubb}},
  \bibinfo {author} {\bibfnamefont {P.~A.}\ \bibnamefont {Coles}}, \bibinfo
  {author} {\bibfnamefont {S.}~\bibnamefont {Diamantopoulou}}, \bibinfo
  {author} {\bibfnamefont {M.~N.}\ \bibnamefont {Gorman}}, \bibinfo {author}
  {\bibfnamefont {C.}~\bibnamefont {Hill}}, \bibinfo {author} {\bibfnamefont
  {A.~Z.}\ \bibnamefont {Lam}},  \emph {et~al.},\ }\href@noop {} {\bibfield
  {journal} {\bibinfo  {journal} {J. Mol. Spectrosc.}\ }\textbf {\bibinfo
  {volume} {327}},\ \bibinfo {pages} {73} (\bibinfo {year} {2016})}\BibitemShut
  {NoStop}%
\bibitem [{\citenamefont {Yurchenko}\ \emph {et~al.}(2018)\citenamefont
  {Yurchenko}, \citenamefont {Szab{\'o}}, \citenamefont {Pyatenko},\ and\
  \citenamefont {Tennyson}}]{yurchenko2018exomol}%
  \BibitemOpen
  \bibfield  {author} {\bibinfo {author} {\bibfnamefont {S.~N.}\ \bibnamefont
  {Yurchenko}}, \bibinfo {author} {\bibfnamefont {I.}~\bibnamefont
  {Szab{\'o}}}, \bibinfo {author} {\bibfnamefont {E.}~\bibnamefont {Pyatenko}},
  \ and\ \bibinfo {author} {\bibfnamefont {J.}~\bibnamefont {Tennyson}},\
  }\href@noop {} {\bibfield  {journal} {\bibinfo  {journal} {Mon. Not. R.
  Astron. Soc.}\ }\textbf {\bibinfo {volume} {480}},\ \bibinfo {pages} {3397}
  (\bibinfo {year} {2018})}\BibitemShut {NoStop}%
\bibitem [{\citenamefont {McKemmish}\ \emph {et~al.}(2020)\citenamefont
  {McKemmish}, \citenamefont {Syme}, \citenamefont {Borsovszky}, \citenamefont
  {Yurchenko}, \citenamefont {Tennyson}, \citenamefont {Furtenbacher},\ and\
  \citenamefont {Cs{\'a}sz{\'a}r}}]{mckemmish2020update}%
  \BibitemOpen
  \bibfield  {author} {\bibinfo {author} {\bibfnamefont {L.~K.}\ \bibnamefont
  {McKemmish}}, \bibinfo {author} {\bibfnamefont {A.-M.}\ \bibnamefont {Syme}},
  \bibinfo {author} {\bibfnamefont {J.}~\bibnamefont {Borsovszky}}, \bibinfo
  {author} {\bibfnamefont {S.~N.}\ \bibnamefont {Yurchenko}}, \bibinfo {author}
  {\bibfnamefont {J.}~\bibnamefont {Tennyson}}, \bibinfo {author}
  {\bibfnamefont {T.}~\bibnamefont {Furtenbacher}}, \ and\ \bibinfo {author}
  {\bibfnamefont {A.~G.}\ \bibnamefont {Cs{\'a}sz{\'a}r}},\ }\href@noop {}
  {\bibfield  {journal} {\bibinfo  {journal} {Mon. Notices Royal Astron. Soc.}\
  }\textbf {\bibinfo {volume} {497}},\ \bibinfo {pages} {1081} (\bibinfo {year}
  {2020})}\BibitemShut {NoStop}%
\bibitem [{Exo()}]{ExoMol}%
  \BibitemOpen
  \href {https://www.exomol.com/data/licence/} {\enquote {\bibinfo {title}
  {Exomol, https://www.exomol.com/},}\ }\BibitemShut {NoStop}%
\bibitem [{\citenamefont {Di~Rosa}(2004)}]{di2004laser}%
  \BibitemOpen
  \bibfield  {author} {\bibinfo {author} {\bibfnamefont {M.}~\bibnamefont
  {Di~Rosa}},\ }\href@noop {} {\bibfield  {journal} {\bibinfo  {journal} {Eur.
  Phys. J. D}\ }\textbf {\bibinfo {volume} {31}},\ \bibinfo {pages} {395}
  (\bibinfo {year} {2004})}\BibitemShut {NoStop}%
\bibitem [{\citenamefont {Fitch}\ and\ \citenamefont
  {Tarbutt}(2021)}]{FITCH2021157}%
  \BibitemOpen
  \bibfield  {author} {\bibinfo {author} {\bibfnamefont {N.}~\bibnamefont
  {Fitch}}\ and\ \bibinfo {author} {\bibfnamefont {M.}~\bibnamefont {Tarbutt}}\
  }(\bibinfo  {publisher} {Academic Press},\ \bibinfo {year} {2021})\ pp.\
  \bibinfo {pages} {157--262}\BibitemShut {NoStop}%
\bibitem [{\citenamefont {Tarbutt}\ \emph {et~al.}(2013)\citenamefont
  {Tarbutt}, \citenamefont {Sauer}, \citenamefont {Hudson},\ and\ \citenamefont
  {Hinds}}]{tarbutt2013design}%
  \BibitemOpen
  \bibfield  {author} {\bibinfo {author} {\bibfnamefont {M.}~\bibnamefont
  {Tarbutt}}, \bibinfo {author} {\bibfnamefont {B.}~\bibnamefont {Sauer}},
  \bibinfo {author} {\bibfnamefont {J.}~\bibnamefont {Hudson}}, \ and\ \bibinfo
  {author} {\bibfnamefont {E.}~\bibnamefont {Hinds}},\ }\href@noop {}
  {\bibfield  {journal} {\bibinfo  {journal} {New J. Phys.}\ }\textbf {\bibinfo
  {volume} {15}},\ \bibinfo {pages} {053034} (\bibinfo {year}
  {2013})}\BibitemShut {NoStop}%
\end{thebibliography}%

\end{document}


\title{Supplementary Material for "Laser Cooling Scheme for the Carbon Dimer ($^{12}$C$_2$)"}

\preprint{APS/123-QED}

\author{N.~Bigagli$^1$}
\author{D.~W.~Savin$^2$}
\author{S.~Will$^1$}
\affiliation{%
$^1$Department of Physics, Columbia University, New York, New York 10027, USA
}
\affiliation{%
$^2$Columbia Astrophysics Laboratory, Columbia University, New York, New York 10027, USA
}
\date{\today}

\maketitle


\section{Calculation of branching ratios}

After identifying three excited states for possible cycling schemes as described in the main text, we identified their decay paths using the ExoMol database \cite{tennyson2016exomol, yurchenko2018exomol, mckemmish2020update, ExoMol} and compiled the respective transition energies and Einstein $A$ coefficients, $A_i$. The subscript $i$ identifies each decay path. In order to determine the relative strength and thus the relative importance of the decay paths, we calculate the branching ratios (BRs) via
\begin{equation}
\text{BR}_i = \frac{A_i}{\sum_j A_j},
\end{equation}
where $j$ runs over all decay paths. Branching ratios are such that 
\begin{equation}
    \sum_j A_j = \Gamma,
\end{equation}
where $\Gamma$ is the total decay rate of the excited state, i.e., $\Gamma = \tau^{-1}$, where $\tau$ is the lifetime of the excited state. Figure 3 of the main text plots the decay BRs for all excited states vs. transition wavelength $\lambda_i$. Decay paths with BRs $<10^{-6}$ are deemed negligible for our considerations and omitted from the calculations and the figure. 

\section{Closure of the cycling schemes}

In order to quantify the level of closure of the cycling schemes, we determine how many scatterings can be performed for a given number of repump lasers addressing the decay paths until $10\,\%$ of molecules remain in cycling-states, that is until $90\,\%$ of the molecules are lost to dark states. In addition, we determine the time necessary to complete those scatterings. 

We model photon scattering as a Bernoulli process with a probability $p$ that the molecule will remain within the cycling scheme per photon scattering \cite{di2004laser} and a probability $1-p$ that the molecule will leave. The quantity $p$ is given by 
\begin{equation}
    p = \sum_i \text{BR}_{i},
\end{equation}
where $i$ runs over all driven transitions (the main cooling transition and the addressed repump transitions). The fraction of molecules in bright states after $n$ photon scatterings is given by $p^{n}$. The number of scatterings that retain $10\%$ of the molecules in a bright state is
\begin{equation}
n_{10\%} = \ln(0.1)/\ln(p).
\end{equation}

The time to complete $n_{10\%}$ scatterings is given by 
\begin{equation}
    t_{10\%} = n_{10\%}/R,
\end{equation} 
where $R$ denotes the scattering rate. The time is related to the natural linewidth of the excited state via
\begin{equation}
    R = \Gamma \cdot n_e,
\end{equation}
where $n_e$ is the probability that the molecule occupies the excited state during the photon cycling process \cite{FITCH2021157}. This takes into account saturation effects and the number of transitions addressed and is given by 
\begin{equation}
    n_e = \frac{1}{G + 1 + 2\sum I_{\text{sat},i}/I_i},
\end{equation}
as discussed in Ref.~\cite{tarbutt2013design}. Here, $I_i$ is the intensity of the laser addressing the $i^\text{th}$ transition; $I_{\text{sat},i} = \pi h c \Gamma / 3 \lambda_i^3$ is the associated saturation intensity \cite{FITCH2021157}; $G$ is the number of driven transitions; $h$ is Planck's constant; and $c$ is the speed of light.

We also note that the definition of saturation intensity for a particular transition, $I_{\text{sat},i}$, between the excited state and one of the lower states in a multi-level system (as relevant here) is identical to the saturation intensity for a two-level system with a scattering rate given by the transition's Einstein $A_i$ coefficient, weighted by the inverse probability that such a transition will occur (given by the inverse of its branching ration), that is 
\begin{equation}
    I_{\text{sat},i} = \frac{\pi h c A_i }{3 \lambda_i^3} \cdot BR ^{-1} = \frac{\pi h c A_i }{3 \lambda_i^3} \cdot \frac{\sum_j A_j }{A_i}.
\end{equation}
The weighing by the inverse branching ratio is done so that a transition that is 100 times weaker than the sum of the other ones (i.e., BR = 0.01) will need 100 times more intensity to get saturated.   

\section{Molecule deflection}

For deflection, molecules scatter photons and experience a recoil that leads to an increasing Doppler shift $\delta$ as more photons are scattered. We assume that molecules continue scattering photons until the Doppler shift is equal to 
\begin{equation}
\delta=\Gamma/2\sqrt{1+I / I_{\text{sat}}}.
\end{equation} 
Given that the Doppler shift is $\delta = v \cdot k = v \cdot \frac{2 \pi}{\lambda}$, where $v$ is the molecule velocity and $k$ is the light wave number, the velocity at which the molecule is tuned out of resonance is given by 
\begin{equation}
v_\delta = \frac{\Gamma\lambda\sqrt{1 + I/I_{sat}}}{4\pi}.
\end{equation}

To find the number of scatterings $n_\text{defl}$ that lead to a recoil velocity $v_\delta$, we divide the momentum associated with this velocity by the photon momentum $p = \hbar k $, with $\hbar = h / 2\pi$. If we only consider the main cooling transition, this yields
\begin{equation}
n_\text{defl} = \frac{m_{\text{C}_2}v_\delta}{p},
\end{equation}
where $m_{\text{C}_2}$ is the mass of a carbon dimer. When repump transitions are addressed, we calculate $v_\delta$ using the strongest transition, i.e.~the transition with largest BR. For the photon momentum, we use the mean value
\begin{equation}
    \bar{p} = \sum_j \hbar k_j \cdot \frac{BR_j}{\sum_i BR_i},
\end{equation}
which weighs the photon momentum of each transition $\hbar k_j$ by the probability of a scattering on transition $j$ to happen. Here, $i$ and $j$ run over the addressed transitions. $L_\text{defl}$, the axial distance travelled by the molecules until they leave resonance, is given by the product of the initial velocity, $v_0$, and the time to complete $n_\text{defl}$ scatterings, $t_\text{defl}=n_\mathrm{defl}/R$, so that $L_\text{defl}=v_0\cdot t_\text{defl}$. The deflection angle is calculated via 
\begin{equation}
    \theta = \arctan(\frac{v_\delta}{v_0}).
\end{equation}

\section{Cooling}

For cooling, we find the number of scatterings $n_\mathrm{cool}$ by dividing the initial molecule momentum by the mean photon momentum $\bar{p}$. Then, we determine the cooling time $t_\text{cool}$ from the scattering rate and the number of photon scatterings via $t_\text{cool} = n_\mathrm{cool}/R$ and the length required for cooling from $L_\text{cool} = v_0\cdot t_\text{cool}/2$. The reported accelerations are calculated via $a = \frac{\hbar k R}{m_{\text{C}_2}}$.

\bibliography{Literature}